\documentclass[prd,twocolumn,amsmath,amssymb,floatfix,superscriptaddress,nofootinbib,preprintnumbers]{revtex4-2}
\pdfoutput=1
\usepackage{color}
\usepackage{graphicx}
\usepackage{tabularx}
\usepackage[dvipsnames, table]{xcolor}
\usepackage{xspace}
\usepackage[justification=centerlast, format=plain, labelfont=bf]{caption}
\usepackage{subcaption}
\usepackage{float}
\usepackage{verbatim}
\usepackage{amsmath}\usepackage{graphicx}
\usepackage{tabularx}
\usepackage{amssymb}
\usepackage{url}
\usepackage{bbold}
\usepackage{slashed}
\usepackage{array}
\usepackage{multirow}
\usepackage{threeparttable}
\usepackage{paralist}
\usepackage{xspace}
\usepackage{upgreek}
\usepackage{lipsum}

\usepackage[colorlinks]{hyperref}
\newcolumntype{C}{>{\centering\arraybackslash}X}

    \newcommand{\be}{\begin{equation}}
  \newcommand{\ee}{\end{equation}}
    \newcommand{\ba}{\begin{align}}
  \newcommand{\ea}{\end{align}}

\newcommand{\Msun}{M_{\odot}}

\begin{document}

\title{Cosmological Constraints on Light (but Massive) Relics}
\author{Weishuang Linda Xu}
\email{weishuangxu@g.harvard.edu}
\affiliation{Department of Physics, Harvard University, Cambridge, MA 02138, USA}
\author{Julian B.~Mu\~noz}
\email{julianmunoz@cfa.harvard.edu}
\affiliation{Harvard-Smithsonian Center for Astrophysics, Cambridge, MA 02138, USA}
\author{Cora Dvorkin}
\email{cdvorkin@g.harvard.edu}
\affiliation{Department of Physics, Harvard University, Cambridge, MA 02138, USA}

\begin{abstract}
Many scenarios of physics beyond the standard model predict new light, weakly coupled degrees of freedom, populated in the early universe and remaining as cosmic relics today. Due to their high abundances, these relics can significantly affect the evolution of the universe. For instance, massless relics produce a shift $\Delta N_{\rm eff}$ to the cosmic expectation of the effective number of active neutrinos. Massive relics, on the other hand, additionally become part of the cosmological dark matter in the later universe, though their light nature allows them to freely stream out of potential wells. This produces novel signatures in the large-scale structure (LSS) of the universe, suppressing matter fluctuations at small scales. We present the first general search for such light (but massive) relics (LiMRs) with cosmic microwave background (CMB) and LSS data, scanning the 2D parameter space of their masses $m_X$ and temperatures $T_X^{(0)}$ today. In the conservative minimum-temperature ($T_X^{(0)}=0.91$ K) scenario, we rule out Weyl (and higher-spin) fermions -- such as the gravitino -- with $m_X\geq 2.3$ eV at 95\% C.L., and set analogous limits of $m_X\leq 11, 1.1, 1.6$ eV for scalar, vector, and Dirac-fermion relics. This is the first search for LiMRs with joint CMB, weak-lensing, and full-shape galaxy data; we demonstrate that weak-lensing data is critical for breaking parameter degeneracies, while full-shape information presents a significant boost in constraining power relative to analyses with only baryon acoustic oscillation parameters. Under the combined strength of these datasets,  our constraints are the tightest and most comprehensive to date.
\end{abstract}

\maketitle


\section{Introduction} 
Cosmology presents a unique avenue to interrogate new-physics scenarios, in strong complementarity to more traditional particle experiments. One particular advantage is the ability of cosmological data to search for \textit{feebly interacting} but \textit{cosmologically abundant} species; indeed, precision data from the cosmic microwave background (CMB) in the early universe~\cite{Aghanim:2018eyx}, as well as surveys mapping the large-scale structure (LSS) of the local universe~\cite{Alam:2016hwk,Erben:2012zw}, have constrained a plethora of beyond-the-standard-model (BSM) scenarios that are presently inaccessible to terrestrial experiments, for instance those involving dark-matter or neutrino self-interactions~\cite{Spergel:1999mh,Buen-Abad:2015ova,Kamada:2016euw,Park:2019ibn}.

An interesting scenario, and a generic prediction of many BSM models, consists of the presence of new \textit{light relics}: weakly interacting particles that decoupled from the Standard Model (SM) thermal bath in the early universe while still relativistic. The canonical example of light relics are the SM neutrinos, which began decoupling just before the epoch of big-bang nucleosynthesis (BBN), when the temperature of the universe was $T_\gamma\sim 10$ MeV~\cite{Mangano:2005cc,deSalas:2016ztq}.  Other hypothetical light degrees of freedom, such as the gravitino~\cite{Weinberg:1982zq,Giudice:1999am,Feng:2010ij,Hook:2015tra,Hook:2018sai}, dark photons~\cite{Ackerman:mha,Vogel:2013raa}, axions~\cite{Peccei:1977hh,Arvanitaki:2009fg,Marsh:2015xka,Baumann:2016wac}, and sterile neutrinos~\cite{Boyarsky:2009ix,Abazajian:2012ys}, are expected to decouple earlier, carrying information about physics at much higher energies. 
We assume that these new particles are stable and negligibly self-interacting, in which case their comoving number density is frozen while relativistic, and a relatively large cosmological abundance --with number densities $\mathcal{O}(10\%)$ that of photons --  survives until the present day.  These scenarios are then especially suited for cosmological searches.

Light relics are commonly assumed to be massless~\cite{Brust:2013ova,Baumann:2017gkg,Green:2019glg}, in which case their effect can be fully absorbed into a change to the effective number $N_{\rm eff}$ of neutrino species (or to the interacting effective number of degrees of freedom if the relics strongly self interact \cite{Bell:2005dr,Brinckmann:2020bcn}), which parametrizes extra contributions to the cosmic radiative energy budget.  
Light-relic masses, however, can have a significant impact on the LSS of the universe.
For instance, neutrino masses, while small enough to be unresolved by current laboratory experiments  (which find $m_\nu \leq $ 1.1 eV at the 90\% confidence level (CL) \cite{Aker:2019uuj}), are best constrained by cosmological datasets.
Similarly, massive light relics will leave a cosmological imprint beyond a background contribution to $N_{\rm eff}$, and will modify the growth of structure at the perturbation level~\cite{Boyarsky:2008xj,Munoz:2018ajr,DePorzio:2020wcz}.  

In this work we conduct a systematic search for Light (but Massive) Relics (LiMRs) with the latest cosmological datasets. We employ data from the CMB, as well as weak-lensing and galaxy surveys, and show that relics with masses $m_X\gtrsim 0.1$ eV are significantly better constrained than their  massless counterparts. 
We rule out minimally coupled Weyl fermions with $m_X>2.3$ eV, real scalars  with $m_X > 11$ eV,  vectors with $m_X>1.6$ eV, and Dirac fermions $m_X>1.1$ eV (all at 95\% CL). These constraints are in fact further applicable to higher-spin fermions and bosons, as only a subset of their available degrees of freedom are active in the early universe due to their ultra-relativistic nature at decoupling; as an example, the $s=3/2$ light relic gravitino is cosmologically equivalent to a $s=1/2$ Weyl fermion~\cite{Vogel:2013raa}. These are the tightest and most comprehensive  constraints on LiMRs to date, and are the first with joint CMB and full-shape galaxy data.

\section{Effect of Relics on CMB and LSS} 
We begin by briefly reviewing the physics of LiMRs and their effect on both the CMB and the LSS of the universe. The reader may refer to e.g.~\cite{Munoz:2018ajr,DePorzio:2020wcz} and Appendix~\ref{AppA} for further details. 

Since light relics decouple from the SM bath while relativistic, they retain their original phase-space distribution, and their temperature thus scales linearly as  $T_X(z) = T_X^{(0)}(1+z)$ with redshift $z$. Their present-day temperature $T_X^{(0)}$ is generically distinct from---and colder than---that of the CMB photons $T_\gamma^{(0)}\approx 2.73$ K, and the difference is determined by when the LiMR decoupled, as after that time the annihilation and decay of unstable SM degrees of freedom heat up only the SM and not the LiMR. In the case of neutrinos, for example, the annihilation of electrons and positrons took place after their decoupling, setting their temperature today at  $T_\nu^{(0)} \approx 1.95$ K~\cite{Mangano:2005cc}. Relics that decouple earlier ought to be colder; accounting for all the known degrees of freedom of the SM sets a minimum temperature of $T_X^{(0)} \approx 0.91$ K for a minimal-extension relic, that is in the case of $m_{\rm new} > T_X^{\rm dec} > m_{\rm top}$ for all non-relic new particles with mass $m_{\rm new}$.  

In the massless limit, the cosmological impact of relics is encapsulated in $\Delta N_{\rm eff} \propto g_X (T_X^{(0)})^4$, which parametrizes the contribution to the radiation energy of a relic with $g_X$ degrees of freedom.  
Its primary effect is on the expansion rate of the universe, which in turn affects the CMB damping tail as well as the phase of the baryon acoustic oscillations (BAOs)~\cite{Bashinsky:2003tk,Baumann:2017gkg}.  The minimum-temperature scenario produces a $\Delta N_{\rm eff}=0.027$ for a massless real scalar (with $g_X=1$ bosonic degree of freedom) or $\Delta N_{\rm eff}=0.047$ for a massless Weyl fermion ($g_X=2$, but fermionic), which sets the targets of future CMB experiments \cite{Ade:2018sbj,Abazajian:2019eic}.


Massive LiMRs, on the other hand, transition to be non-relativistic when $T_X(z)$ becomes comparable to their mass $m_X$, no longer contributing to the energy budget of the universe as radiation.
After this epoch, LiMRs behave as a subcomponent of the cosmological dark matter (DM). Unlike the majority of DM which is cold (referred to as CDM), LiMRs have significant streaming motions due to their temperature, which impedes their clustering beyond a characteristic free-streaming scale~\cite{Lesgourgues:2006nd}. Therefore, LiMRs (like neutrinos) behave as a type of hot DM, impacting the growth of matter fluctuations and thus the observable LSS of the universe.\footnote{LiMRs could also exhibit clustering, depending on their mass, but we will not include this effect here as it is poorly understood beyond the case of neutrinos~\cite{LoVerde:2013lta}.}

\begin{figure}
    \centering
    \includegraphics[width=\linewidth]{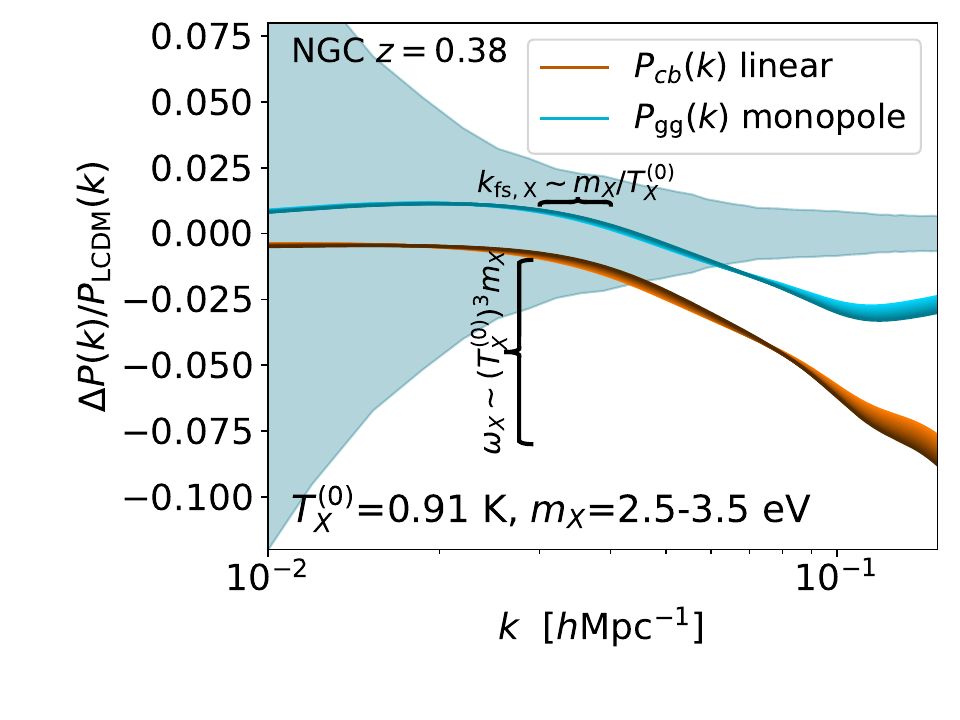}
    \caption{Effect of LiMRs on the clustering of matter and galaxies. 
    The orange lines represent the change to the real-space linear matter power spectrum due to LiMRs with different masses,
    whereas the blue lines show the nonlinear redshift-space galaxy power spectrum, as observed by the SDSS, where the shaded region represents 1$\sigma$ errors on the BOSS DR12 NGC patch~\cite{Ivanov:2019pdj}, all at $z= 0.38$.
    Both power spectra are compared to a $\Lambda$CDM baseline cosmology where the parameters $\{\omega_b, h, n_s, A_s, \sum m_\nu \}$ are held fixed, while $\omega_{\rm cdm}$ is adjusted to compensate for a range of smaller (lighter) and larger (darker) relic masses.
    The ability to independently measure the scale ($\propto k_{\rm fs}$) and amplitude ($\propto \omega_X$) of the LiMR-induced suppression allows us to reconstruct both the mass and temperature of any LiMR.}
    \label{fig:limr_nonlinear}
\end{figure}

The effect of LiMRs on the LSS is illustrated in Fig.~\ref{fig:limr_nonlinear}, where we show the power spectrum (i.e., the Fourier-space two-point function) of  matter (CDM + baryon) fluctuations at the linear level, and see a scale-dependent suppression of power.
The scale of this suppression is set by the relic free-streaming~\cite{Ali-Haimoud:2012fzp}
\begin{equation}
k_{\rm fs} = \frac{ 0.8\, h\text{ Mpc}^{-1} }{\sqrt{1+z}} \left( \frac{m_\chi}{ 1  \text{ eV}}\right) \left(\frac{T_X^{(0)}}{ T_\nu^{(0)} }\right)^{-1},
\end{equation}
and its amplitude is roughly $ 14\, \omega_X/\omega_m$~\cite{Lesgourgues:2006nd}, given the present-day relic abundance
\begin{equation}
\omega_X = 0.011 \left( \frac{g_X}{g_\nu} \right) \left(\frac{m_X}{1 \text{ eV}}\right) \left(\frac{T_X^{(0)}}{ T_\nu^{(0)} }\right)^{3}.  
\end{equation}
Thus, by studying the location and depth of this suppression, one can disentangle the LiMR mass and temperature, as well as distinguish the presence of new LiMRs from the background of SM neutrinos. In practice, however, care must be taken to include quasi-linear corrections, as well as the biasing of the observable galaxies, which are imperfect tracers of the underlying matter fluctuations.
We account for these effects using the publicly available {\tt CLASS-PT}~\cite{Chudaykin:2020aoj}, an extension to the {\tt CLASS} Boltzmann code~\cite{2011JCAP...07..034B}  which incorporates much of the theoretical progress on understanding the LSS of the universe, including biasing and perturbation theory to the one-loop level. 
These effects slightly modify the predictions for the power spectrum in the presence of a LiMR, especially at higher $k$, as shown in Fig.~\ref{fig:limr_nonlinear} and discussed in further detail in Appendix~\ref{AppA}. We assume that, in the presence of massive relics, observed galaxies are biased only with respect to the subset of clustering matter (that is, CDM + baryons), as is the case for neutrinos~\cite{Villaescusa-Navarro:2013pva,Biagetti:2014pha,LoVerde:2014pxa}. Additionally, we note that {\tt CLASS-PT} incorporates the effect of massive neutrinos and light relics at linear order only, and a more precise perturbative treatment has yet to be developed. We expect this precision to be sufficient for currently accessible sensitivities, as the next-order corrections, of order $(\omega_X/\omega_m)^2 \delta_{\rm cb}^2 \lesssim 10^{-3}$, are presently beyond experimental reach.


\section{The data and likelihoods} 
We now describe the datasets we employ to search for LiMRs.
For the CMB, we use the full TT, TE, EE, low-E, and lensing likelihoods from the {\it Planck} 2018 public release~\cite{Aghanim:2018eyx}, encapsulating measurements of the temperature and polarization anisotropies as well as lensing information, which we will collectively call {\tt P18}.  We add to this weak-lensing data from the Canada-France-Hawaii Telescope (CFHTLens) collaboration~\cite{Heymans:2013fya}, consisting of 2-point correlation functions of galaxy ellipticities for which we follow the prescription for a ``conservative" tomographic analysis, and we denote this as {\tt WLens}.  Finally, we take galaxy power-spectrum data from BOSS data release 12~\cite{Alam:2016hwk}, which contains spectroscopic information of $\mathcal{O}(10^6)$ galaxies over two redshift bins (with $z_{\rm eff} =0.38$ and 0.61), each split into two spatial subsets based on Galactic hemisphere (termed the North/South Galactic Cap, or N/SGC)~\cite{Ivanov:2019pdj}.  
We denote this likelihood as {\tt BOSS-FS}, and stop our analysis at scales of $k_{\rm max} =0.25h/\text{Mpc}$, where the perturbative approach is still valid.  We will assess the relative advantage of incorporating full-shape LSS data by also considering only the BAO-based standard-ruler parameter information obtained by the BOSS survey (referred to as {\tt BOSS-BAO})~\cite{Alam:2016hwk}. We marginalize over the full set of beam and foreground nuisance parameters associated with the {\tt P18} likelihoods~\cite{Planck:2019nip}. For the {\tt BOSS-FS} likelihood we vary three nuisance parameters $\{b_1, b_2, b_{\mathcal{G}_2}\}$ for each of the four datasets, corresponding to the linear, quadratic, and tidal biases respectively; several additional parameters, such as counterterms, are marginalized over internally within the likelihood~\cite{Philcox:2020vvt}. For the weak lensing dataset, we omit marginalization over galaxy intrisic alignment, as no significant signal for such has been found in the late-type galaxies used in the data~\cite{Heymans:2013fya}.

Before searching for any LiMR,  we note that there is a well-known (though not highly statistically significant~\cite{Lemos:2020jry,Abbott:2021bzy}) discrepancy in the amplitude of clustering between measurements from the CMB and from late-universe lensing/LSS data, commonly termed the ``$\sigma_8$ tension". This might have ostensibly complicated the validity of combining these datasets for a joint analysis, but we find in fact that the effect of LiMRs is largely orthogonal to the axis of disagreement, and while our constraints are significantly improved by the combined power of all three datasets, the addition of LiMRs does little to ameliorate or exacerbate this tension. 
This is because while the incorporation of cosmological LiMRs does induce a small shift in the inferred value of $\sigma_8$, it does so at the expense of displacing some CDM abundance into LiMRs (i.e., reducing the CDM abundance $\omega_{\rm cdm}$ to keep $\omega_m$ constant).
This is strongly disfavored by weak-lensing data, which constrains $\omega_{\rm cdm}$ particularly well.  
A more detailed discussion on the interplay of LiMRs and the $\sigma_8$ tension can be found in Appendix~\ref{AppB}.

\section{Full Parameter Space} 
A cosmological population of stable, non-self interacting LiMRs is fully described by their mass $m_X$, present-day temperature $T^{(0)}_X$, and fermionic or bosonic degrees of freedom $g_X$.
It was shown in Ref.~\cite{Munoz:2018ajr} that in fact, a traversal of just two of the three parameters is sufficient to search for evidence of LiMRs of any type, Bosonic or Fermionic, fully spanning the space of observable signatures. Details of this direction of degeneracy and the translation between relics with equivalent cosmological imprint can be found in Appendix~\ref{AppB}.  In our analysis, we sweep through $\{m_X, T_X^{(0)}\}$ for a (neutrino-like) Weyl fermion to search for all types of LiMRs. We assume linear-flat priors on both parameters, imposing additionally a hard prior on relic masses $m_X \in [0, 20]$ eV. In addition to these two new physics parameters, we marginalize over the standard cosmological parameters $\{\omega_b, \omega_{\rm cdm}, h , \tau_{\rm reio}, n_s, A_s\}$, 
where $\omega_b$ is the baryon abundance, $h$ is the reduced Hubble parameter, $A_s$ and $n_s$ parametrize the primordial power spectrum, and $\tau_{\rm reio}$ is the optical depth to reionization. Finally, we also marginalize over neutrino masses $\sum m_\nu$, assuming 3 degenerate massive neutrinos. Our cosmological model is then fully described by these 9 parameters.

\begin{figure}
    \centering
    \includegraphics[width=\linewidth]{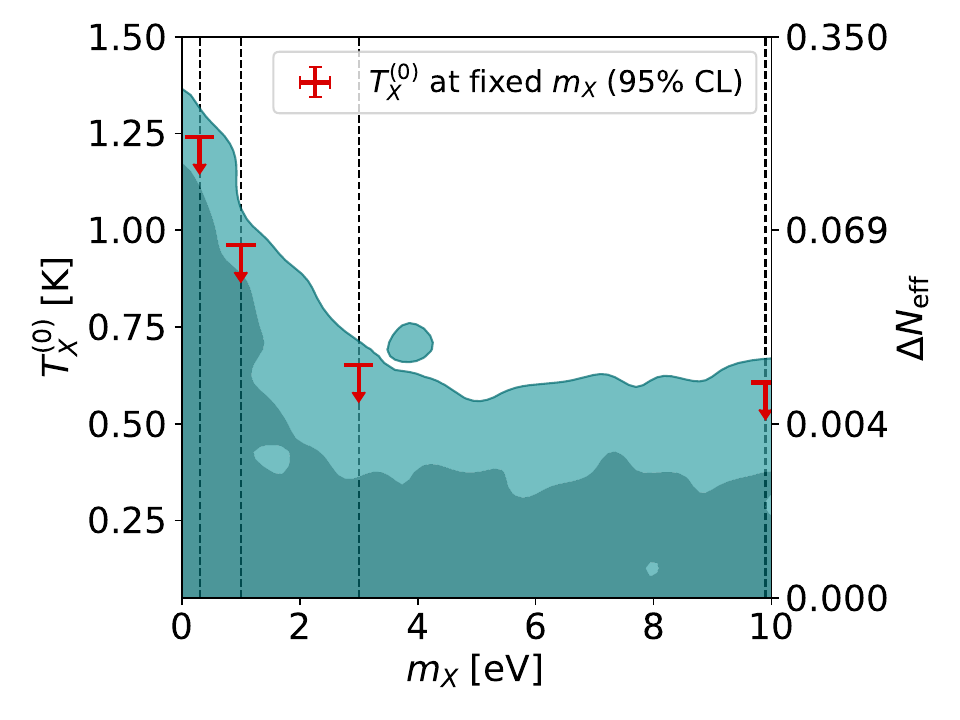}
    \caption{ Constraints for Weyl-fermion light relics in the $m_X - T_X^{(0)}$ parameter space, obtained from a joint analysis of the {\tt P18 + BOSS-FS + WLens} datasets, where 68\% (95\%) CL limits are shown as dark (light) blue. 
    We find no preference for relics throughout, and display specific 95\% CL upper bounds on present-day temperature for relics of fixed mass as vertical red arrows. 
    Other types of relics (such as scalars or vectors) have identical signatures to Weyl fermions with different parameters, so this search rules out LiMRs of any spin.
    }
    \label{fig:2D_constraints}
\end{figure}

We perform a Markov Chain Monte Carlo (MCMC) likelihood analysis with the Metropolis-Hasting algorithm to search through the full LiMR parameter space, by applying a conservative Gelman- Rubin criterion of $R-1 < 0.01$. We show our results in Fig.~\ref{fig:2D_constraints}, displayed along the basis of mass and temperature of a Weyl fermion; we do not find evidence for a cosmological population of LiMRs.
The dark (light) contours show the region of parameter space allowed by the data ({\tt P18 + BOSS-FS + WLens}) at 68\% (95\%) CL. Note that the limit of low temperature corresponds to a relic species with vanishing number and energy abundance, which cannot be constrained. High-mass ($m_X\gtrsim 10 eV$) LiMRs can have sizeable abundances even for low $T_X$, though these relics have $k_{\rm fs}$ larger than the wavenumbers we can probe, and behave effectively as CDM. 

In order to find precise constraints, we perform 1D scans for LiMRs of fixed masses $m_X=\{0.01,0.03, 0.1, 0.3, 1, 3,10\}$ eV, and find that $T_X^{(0)} < \{1.3, 1.3, 1.3, 1.2, 0.96, 0.65, 0.61 \}$ K at 95\% CL, respectively, shown as vertical arrows in Fig.~\ref{fig:2D_constraints}. 
This illustrates that LiMRs are easier to detect if they have higher masses, though for $m_X\gtrsim 3$ eV our constraints plateau as sufficiently cold relics become indistinguishable from cold DM independent of mass. 

\section{Minimum-Temperature Relics} 
As advanced above, there is a minimum temperature ($T_X^{(0)}=0.91$ K) that a minimal-extension relic can have,
reached only for relics that decouple while all of the SM degrees of freedom were still cosmologically populated. 
This scenario represents the most conservative contribution that each species of LiMR can yield. 
We will thus perform a 1D scan varying the LiMR mass $m_X$, but keeping $T_X^{(0)}=0.91$ K, for each of the four types of relic we consider: scalar, Weyl, vector, and Dirac. We obtain upper limits on particle mass of $11$ eV for scalars, $2.3$ eV for Weyl fermions, $1.6$ eV for vectors, and $1.1$ eV for Dirac fermions, all at 95\% CL. 

\begin{figure}
    \centering
    \includegraphics[width=\linewidth]{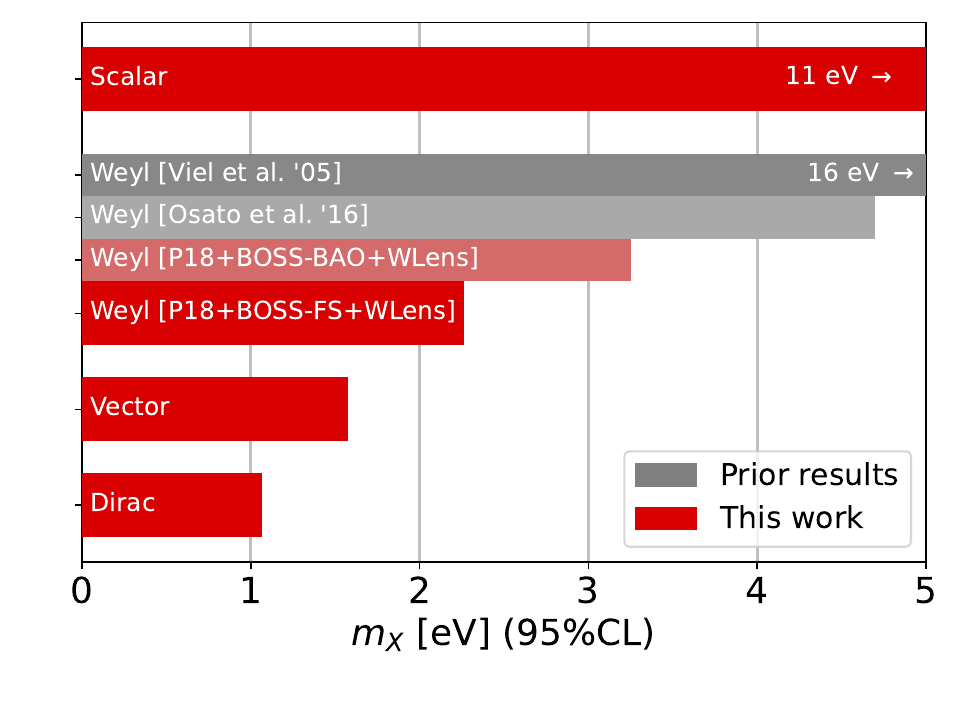}
    \caption{Limits on the mass $m_X$ of different species of light relic, all at 95\% CL and assuming the minimum-temperature scenario, $T_X^{0}= 0.91$ K. 
    Red bars show constraints from this work, which are obtained via joint analysis of all our data sets ({\tt P18+BOSS-FS+WLens}), whereas the pink band has BAO-only rather than full-shape galaxy data. 
    Gray bands represent the previous constraints on Weyl fermions from Refs.~\cite{Viel:2005qj,Osato:2016ixc}.
    Our limits are a factor of $2-5$ stronger and extend to other relic species.
   } 
    \label{fig:bargraph}
\end{figure}

We summarize these constraints in Fig.~\ref{fig:bargraph}, and compare to previous work.
The previous limits were limited to Weyl relics, and used CMB data added to the Lyman-$\alpha$ forest~\cite{Viel:2005qj} or BAO and weak-lensing data~\cite{Osato:2016ixc}\footnote{We note that Ref.~\cite{Osato:2016ixc} assumes a slightly higher relic temperature, which is less conservative. We recover excellent agreement with that work under matching assumptions.}) and our constraints are stronger by a factor $2-5$. 
We also investigate here the relative power of each dataset, and find 
({\it i\,}) the inclusion of full-shape galaxy power-spectrum information, as opposed to BAO only, strengthens our constraints by a significant $30\%$, 
and ({\it ii\,})  weak-lensing data is crucial for obtaining strong limits, as it precisely measures the abundance of clustering matter, breaking a degeneracy between $\omega_{\rm cdm}$ and $\omega_X$ (the reader can find the confidence contours in Appendix~\ref{AppB}).

Our constraints on LiMRs can be interpreted within different particle-physics models: eV-scale extensions to the neutrino sector, particularly sterile neutrinos, have been widely proposed and studied~\cite{Boyarsky:2009ix,Abazajian:2012ys,Chacko:2016hvu}, dark photons~\cite{Ackerman:mha,Vogel:2013raa,Cyr-Racine:2012tfp} are well-motivated examples of a vector LiMR, and scalar relics are straightforwardly realized in axions and axion-like particles~\cite{Peccei:1977hh,Arvanitaki:2009fg,Marsh:2015xka,Baumann:2016wac}. We note, however, in the latter case that our present data is insensitive to the sub-eV mass candidates typically considered, though a relic population of hot QCD axions are expected to have much higher than minimum temperature~\cite{DEramo:2018vss}.

As a detailed example, we study the case of the gravitino, for which a relic population easily arises in gauge-mediated SUSY-breaking scenarios~\cite{Dine:1993yw,Dine:1994vc,Giudice:1998bp}. While the gravitino is intrinsically $s=3/2$,
only two of its four modes are thermally populated at the time of its relativistic decoupling, making it cosmologically equivalent to a Weyl fermion ($s=1/2$), and allowing us to set a limit on its mass $m_X<2.3$ eV at 95\% CL.
This limit is strictly conservative, as the gravitino decoupling temperature can only be higher than our minimal $T^{(0)}_X = 0.91$ K for these models~\cite{Ichikawa:2009ir}. 
Our limit cuts into the predictions of low-energy SUSY-breaking scenarios~\cite{Hook:2015tra,Hook:2018sai}.
Consequently, we are able to set an upper limit on the SUSY breaking scale, estimated as $\sqrt{F} \approx \sqrt{M_{\rm pl} m_X} \leq 70$ TeV~\cite{Dine:1993yw,Giudice:1998bp,Osato:2016ixc}, where $M_{\rm pl}$ is the reduced Planck mass, in strong complementarity with upcoming lower bounds from collider studies~\cite{FCC:2018evy,Schulte:2017qkc,Hinchliffe:2015qma}. 


\section{Discussion and conclusions} 
In this work we present the strongest constraints to date on cosmological light (but massive) relics, and the first ever to make use of full-shape LSS data. The inclusion of broadband galaxy data as well as state-of-the-art CMB measurements allows us to improve significantly upon previous limits, and to present comprehensive bounds across the parameter space of relics of various species, masses, and temperatures.  We find that low-redshift weak-lensing data is critical to break key degeneracies, and the orthogonality of the LiMR signature with the $\sigma_8$ tension allows us to safely incorporate those data.

The coming years will see a dramatic improvement in the amount of cosmological data available, as new CMB facilities and galaxy surveys will come online.
These data have the potential to unearth any LiMRs that populate our universe.
We have shown that LiMRs with $m_X\gtrsim 0.1$ eV leave a distinct imprint in the LSS, and are easier to detect than their massless counterparts.
Thus, it is possible that the first relic discovered is massive.
This would open a window to the state of the cosmos at energies beyond those presently accessible by colliders, showing the promise of cosmological data sets to understand high-energy physics.


\begin{acknowledgments}
We thank Prateek Agrawal,  Emanuele Castorina, Mikhail Ivanov, Simon Knapen, Oliver Philcox and Marko Simonovi\'c for thoughtful comments on an earlier draft of this work. We thank Nicholas Deporzio for helpful discussions and collaboration in the early stages of this work. We acknowledge the use of \texttt{MontePython 3.3} and \texttt{CLASS-PT} for the analysis conducted in this work. WLX was partially supported by the Moore Foundation Award 8342. JBM was funded by a Clay Fellowship at the Smithsonian Astrophysical Observatory. 
CD was partially supported by the Department of Energy (DOE) Grant No. DE-SC0020223. 
\end{acknowledgments}

\clearpage
\appendix

\section{Effect of LiMRs on CLASS-PT}
\label{AppA}

Our ability to constrain cosmological parameters, including the presence of cosmological LiMRs (or lack thereof), is dependent on our theoretical control of the model predictions. To accurately model the galaxy power spectra in the quasi-linear regime---where relic effects become important---we incorporate perturbative corrections to the linear model at the one-loop level. In particular, dedicated studies on the impact of massive neutrinos on the EFT framework have been presented in Refs.~\cite{Angulo:2015eqa,Senatore:2017hyk}.  In this work we follow the framework of Refs.~\cite{Chudaykin:2020aoj,Ivanov:2020ril} and encourage the reader to read these works for further details. As mentioned in the main text, we emphasize here that in our present treatment the effects of massive neutrinos and relics are incorporated up to first order, and nonlinear effects such as dynamic interactions between relic and CDM fluids have been neglected. The development of a complete perturbative prescription for the cosmology of new massive particles, as well as its use to constrain LiMRs at high precision, is reserved for future work.

In harmonic space, the 2D-anisotropic redshift-space galaxy power spectra can be modeled as
\begin{equation}
    P_{\ell} = P^{\rm lin}_\ell + P^{\rm 1-loop}_\ell + P^{\rm ctr}_\ell  + P^{\rm noise}_{\ell},
\end{equation}
where each constituent term ($P_\ell^X$ for linear, 1-loop, counterterm and noise contributions) is related to its real-space matter counterpart by bias and redshift-space distortion terms. More pertinently, the galaxy-density perturbation $\delta_g$ is related to its matter counterpart $\delta$ via the Taylor expansion
\begin{equation}
    \delta_g \equiv b_1 \delta + b_2 \delta^2 + b_{\mathcal{G}_2}\mathcal{G}_2, 
    \label{eq:deltag1loop}
\end{equation}
where $\mathcal{G}_2$ is the tidal operator, with convolved dependencies on the linear matter perturbations,
and $\{b_1, b_2, b_{\mathcal{G}_2}\}$ are biases that we vary as nuisance parameters, independently for each of the four datasets (the North/South Galactic Caps and two redshift bins $z_{\rm eff}=0.38$ and 0.61).
An additional bias parameter $b_{\Gamma_3}$ was found to be degenerate under the sensitivity level of BOSS data and is held fixed in this analysis, and the counterterms $\{c_0, c_2, \tilde c\}$ are internally marginalized over within the likelihood~\cite{Ivanov:2019pdj,Philcox:2020vvt}.

The incorporation of massive light relics into the cosmology splits the total matter abundance into clustering (CDM+baryon) and non-clustering (neutrino and LiMR) constituents, where galaxies are biased tracers of only the former \cite{Villaescusa-Navarro:2013pva,Castorina:2013wga}. 
Nevertheless, the free-streaming of light relics at small scales suppresses also the growth of CDM+baryon fluctuations~\cite{Lesgourgues:2006nd,Green:2019glg}. As laid out in the main text, this effect on structure formation can be described at linear order with a characteristic scale ($k_{\rm fs}$) and amplitude ($\omega_X$). 

\begin{figure*}
    \centering
    \includegraphics[width=0.7\linewidth]{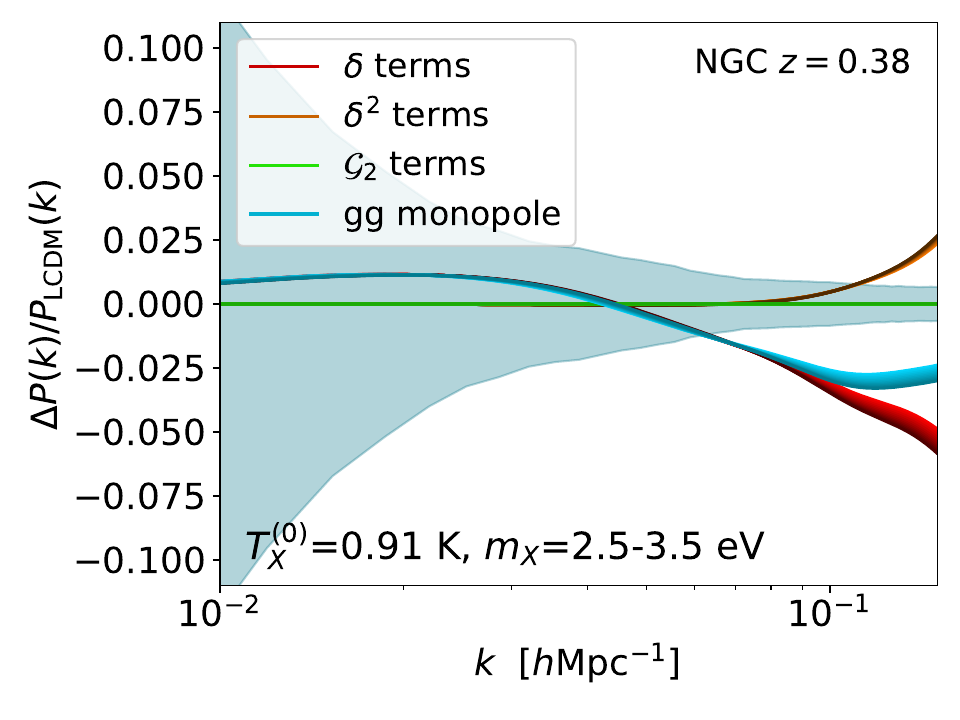}
    \caption{The effect of LiMRs on the redshift-space galaxy fluctuation monopole (blue), and the breakdown of this effect into constituent terms that depend on only the linear matter fluctuations (red), on its quadratic counterpart (orange), and on the tidal operator (green), all as defined in Eq.~\eqref{eq:deltag1loop}. The turn-around of the LiMR-induced small-scale suppression of the monopole is driven by the higher-order corrections whose own LiMR-induced suppression is negatively biased. In this figure the parameters are fixed and evaluated as in Fig.~\ref{fig:limr_nonlinear}
    }
    \label{fig:limr_breakdown_nonlinear}
\end{figure*}

In this quasi-linear framework, however, the presence of LiMRs suppresses not only the linear contribution ($b_1 \delta$) to the galaxy perturbations $\delta_g$ at small scales, but does so for each of the higher-order terms in Eq.~\eqref{eq:deltag1loop} as well. 
Therefore, the sign of each of the biases ($b_2, b_{\mathcal G_2}$) is critical for determining whether the magnitude of the suppression is overall strengthened or diminished by these one-loop terms.  For BOSS galaxy haloes, with $M\sim 10^{12} \Msun$ at $z\sim 0.5$, the bias parameters are expected to be~\cite{Ivanov:2019pdj,Lazeyras:2015lgp}
\begin{equation}
    b_1 \sim 2 \qquad b_2 \sim - 0.6 \qquad b_{\mathcal{G}_2} \sim -0.3,
\end{equation}
where the relevant feature is that the quadratic bias is negative.
 We show the mean and $1-\sigma$ widths for the bias nuisance parameters recovered from our analysis in Tab.~\ref{tab:bias_nuisance}.

\begin{table*}[]
    \centering
\begin{tabular}{c | c  | c  | c | c }
\hline 
\hline
 & \multicolumn{2}{c}{\qquad North Galactic Cap \qquad} & \multicolumn{2}{c}{\qquad South Galactic Cap \qquad} \\
& \multicolumn{1}{c}{z=0.38} &\multicolumn{1}{c}{z=0.61} &\multicolumn{1}{c}{z=0.38}&\multicolumn{1}{c}{z=0.61}\\
\hline
\hline
 $b_1$ & $\quad 1.96 \pm 0.053\quad$ &  $\quad2.04 \pm 0.067\quad$ & $\quad1.86 \pm 0.051\quad$ &  $\quad1.87 \pm 0.065\quad$  \\
 $b_2$ & $\quad -2.26 \pm 0.66\quad$  & $\quad-1.24 \pm 0.78\quad$ & $\quad-1.46 \pm 0.54\quad$ & $\quad-1.75 \pm 0.57\quad$\\
 $b_{\mathcal{G}_2}$ &  $\quad -0.17 \pm 0.25\quad$ & $\quad0.06 \pm 0.28\quad$ & $\quad-0.19 \pm 0.16\quad$ & $\quad0.21 \pm 0.24\quad$\\
 \hline \hline 
\end{tabular}
    \caption{ Mean and widths of the bias posteriors recovered from the {\tt BOSS-FS} dataset, assuming a cosmology with a $T_X^{(0)} = 0.91$ K Weyl fermion LiMR. Linear, quadratic, and tidal biases are varied as nuisance parameters, independently for each of the four datasets. } 
    \label{tab:bias_nuisance}
\end{table*}

In that case, a portion of the original suppression is canceled out, and indeed the resultant galaxy power spectra is predicted to turn-around with a mild enhancement at the smallest scales; this effect is illustrated in Fig.~\ref{fig:limr_breakdown_nonlinear}.
This understanding of how the presence of LiMRs affects nonlinear power spectra suggests a strategy for future searches of these new particles. 
Measurements of halo biases in simulations~\cite{Lazeyras:2015lgp} have recovered positive quadratic bias for more-massive haloes of $M \gtrsim 10^{14} M_\odot$, so for galaxies hosted in these haloes we should expect an enhanced sensitivity to light relic masses, as the small-scale suppression from both linear and higher-order contributions co-add rather than cancel out.

Finally we note that in addition to the effects described above, LiMRs give rise to a scale-dependent step in the linear galaxy bias $b_1$ with respect to CDM+baryons due to the scale-dependent growth they produce~\cite{Xu:2020fyg,Vagnozzi:2018pwo,Chiang:2018laa}. This growth-induced scale-dependent bias (GISDB) is computationally implemented in {\tt RelicCLASS}\footnote{\url{https://github.com/wlxu/RelicClass}} ( based on {\tt RelicFAST}\footnote{\url{https://github.com/JulianBMunoz/RelicFast}}), though in previous work we showed that the BOSS data are not accurate enough to resolve an effect of this magnitude~\cite{Xu:2020fyg}. We therefore neglect the GISDB effect in this search as well.

\section{Parameters and  Degeneracies in the Minimum-Temperature Case}
\label{AppB}

A species of cosmological LiMRs is fully described by their mass $m_X$, present-day temperature $T^{(0)}_X$, and degrees of freedom $g_X$. These determine the observational signature of a particular LiMR species, quantified as the contribution to $\Delta N_{\rm eff} \propto g_X (T_X^{(0)})^4$, the free-streaming scale $k_{\rm fs}\propto m_X/T_X^{(0)}$, and the matter abundance $\omega_X\propto g_X m_X (T_X^{(0)})^3$. However, since this latter quantity is proportional to the product of the former two, there is a flat direction in this parameter space of relics that are observationally equivalent. Explicitly, a relic of any species described by $\{m_X, T^{(0)}_X, g_X\}$  is cosmologically indistinguishable from a Weyl fermion with  
\begin{equation}
   m_W =  m_X (g_X/g_W)^{1/4} c_1^{\gamma/4} c_2^\gamma \qquad T_W = T^{(0)}_X (g_X/g_W)^{1/4} c_1^{\gamma/4} ,
   \label{eq:equiv_relic}
\end{equation} where $c_1=8/7$, $c_2 = 7/6$, and $\gamma = 0$ or 1 for an originally fermionic or bosonic relic~\cite{Munoz:2018ajr}, and $g_W =2$. This allows us to search through the full space of all possible relics with a sweep of two of the three parameters, as done in Fig.~\ref{fig:2D_constraints}. In the same vein, the limits we present for various species in the minimum-temperature scenario (Fig.~\ref{fig:bargraph}) can be equivalently interpreted as bounds for a single species of relic --  e.g. the neutrino-like Weyl -- at different temperatures: $m_W < \{ 11.2, 2.26, 1.90, 1.27\}$ eV for Weyl relics with $T_W^{(0)} = \{0.79, 0.91, 0.94, 1.08 \}$ K, all at 95\% CL. 

\begin{figure*}
    \centering
    \includegraphics[width=\linewidth]{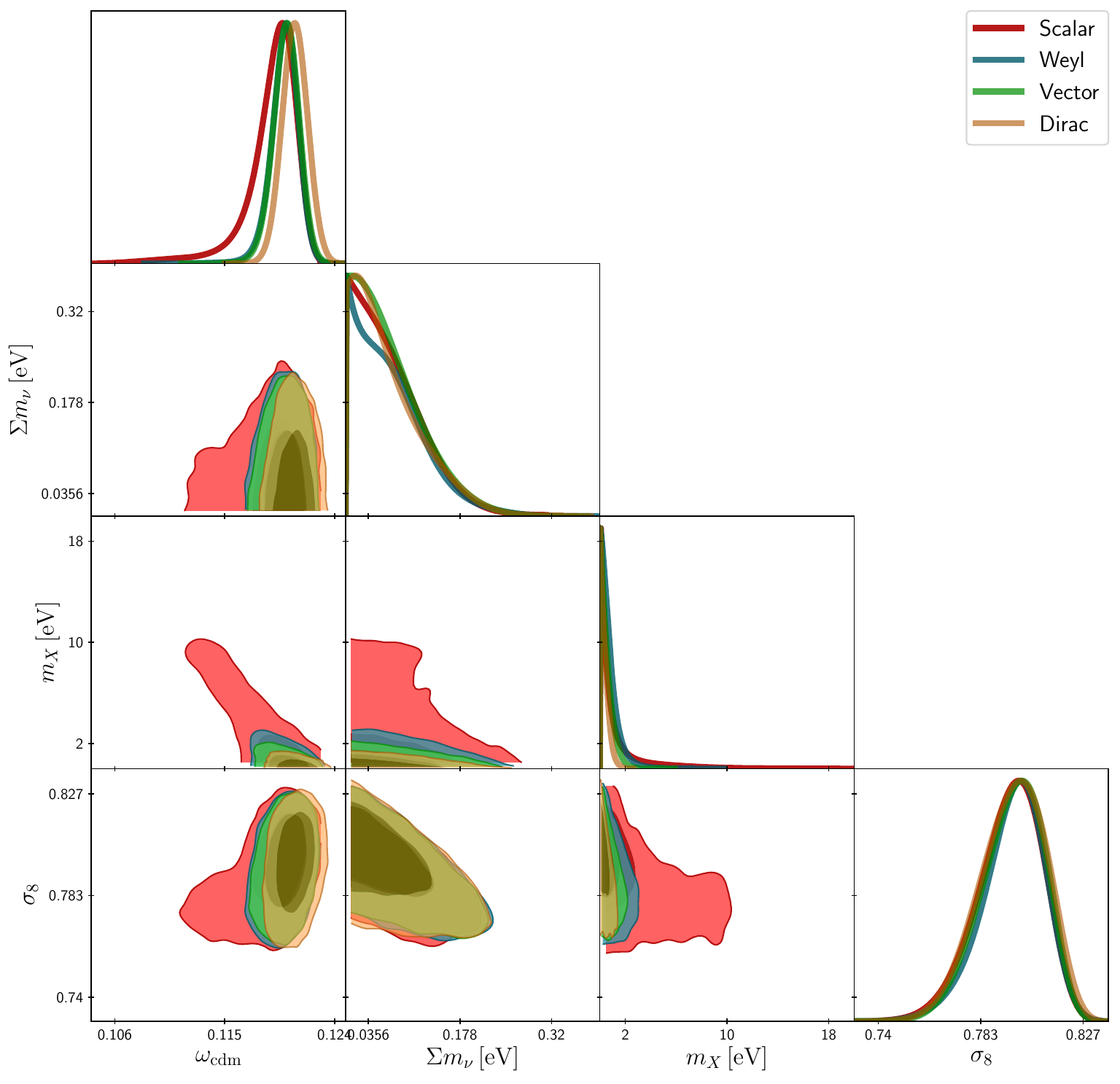}
    \caption{One and two-dimensional probability distribution functions for the subset of cosmological parameters most degenerate with LiMRs (in the $T_X^{(0)} = 0.91$ K minimum-temperature scenario), under a joint analysis of {\tt P18 + BOSS-FS + WLens} data. Particles with fewer degrees of freedom, such as real scalars, have a relatively smaller effect and are allowed to displace a higher fraction of the CDM abundance, making them comparatively difficult to detect.}   \label{fig:Tmin_triangle}
\end{figure*}

In the remainder of this section we will use the minimum-temperature relic scenarios as an example to study how much, if at all, the cosmological effects induced by LiMRs are degenerate or correlated with other cosmological parameters, with the understanding that these observations straightforwardly extend to a range of relic temperatures.  In particular, we will be interested in studying the degeneracy between the LiMR mass $m_X$ (at $T_X^{(0)}=0.91$ K) and the total neutrino mass ($\sum m_\nu$), as well as the abundance $\omega_{\rm cdm}$ of cold (non-LiMR) DM, which is the main component of clustering matter along with baryons. We note that $\omega_b$ can be easily distinguishable from $\omega_{\rm cdm}$ and $\omega_X$ with CMB and BAO measurements, as well as BBN, and therefore CDM is the dominant component that is displaced by the injection of light relics. We also consider $\sigma_8$, as opposed to correlations with $n_s$ or $A_s$ individually, as this quantity most readily quantifies the amplitude of clustering in the large-scale structure. We show the triangle (corner) plot with 2D confidence contours between these key cosmological parameters in Fig.~\ref{fig:Tmin_triangle}, under a joint analysis between CMB ({\tt P18}), LSS ({\tt BOSS-FS}), and weak-lensing ({\tt WLens}) data.

One of the key takeaways from this figure is that neutrino and LiMR masses are not strongly degenerate, and can be distinguished. In particular, we find a limit on the sum of neutrino masses of $\sum m_\nu < 163$ meV at 95\% CL (assuming degenerate neutrino masses). This is comparable to results found in previous work~\cite{Ivanov:2019pdj,Vagnozzi:2017ovm}. In contrast, previous searches for specific massive relics such as the gravitino~\cite{Osato:2016ixc} did not marginalize over $\sum m_\nu$, and instead took the neutrinos to be massless.

Instead, the main degeneracy depicted in Fig.~\ref{fig:Tmin_triangle} is between the LiMR mass $m_X$ and the $\omega_{\rm cdm}-\sigma_8$ combination.  To keep $\omega_m$ fixed, the effect of a larger LiMR mass $m_X$ can be partly compensated by reducing $\omega_{\rm cdm}$. This, however, results in an overall smaller amount of clustering matter, and consequently a lower observed value of $\sigma_8$.  This degeneracy is most readily observable for the scalar case in Fig.~\ref{fig:Tmin_triangle}, as it has the lowest effective temperature (equivalent to a Weyl fermion with $T_X^{(0)}=0.79$ K), and thus it is most similar to, and therefore degenerate with, CDM. 

\begin{figure*}
    \centering
    \includegraphics[width=0.8\linewidth]{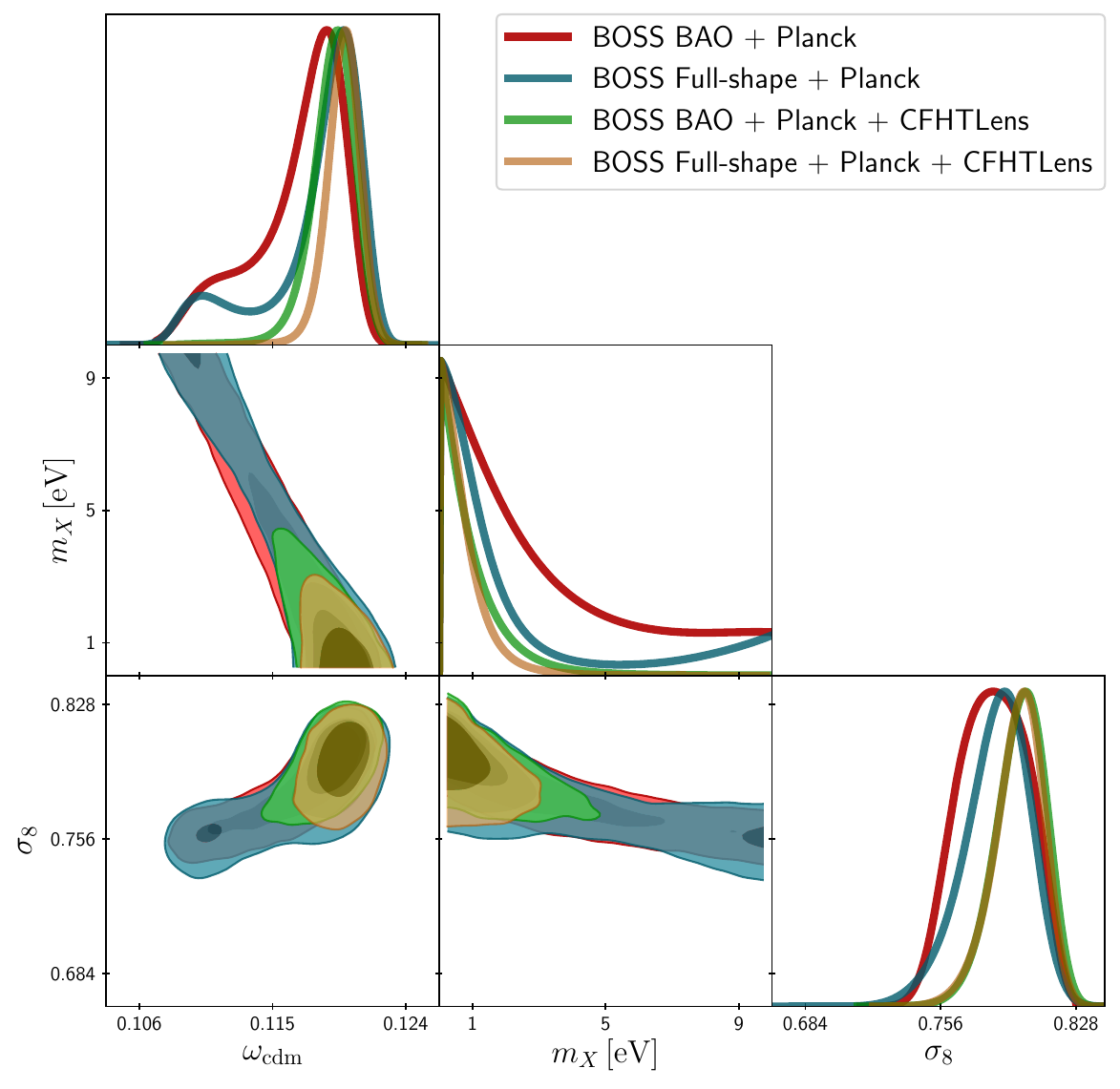}
    \caption{Similar to Fig.~\ref{fig:Tmin_triangle}, but comparing different datasets -- {\tt P18 + BOSS-BAO} (red), {\tt P18 + BOSS-FS} (blue), {\tt P18 + BOSS-BAO + WLens} (green), and {\tt P18, BOSS-FS, Wlens} (yellow) -- all assuming the scenario of a minimum-temperature Weyl relic. The incorporation of CFHTLens data is crucial to establish constraints, as these data measure the abundance of clustering matter, and thus break a strong $m_X - \omega_{\rm cdm}$ degeneracy. 
    While weak-lensing data typically prefers a smaller value of $\sigma_8$ than measured by CMB experiments, it is shown here to disfavor the lowering of this value at the cost of incorporating LiMRs and lowering the abundance of clustering (CDM + baryon) matter.
    We thus conclude that LiMRs cannot solve the $\sigma_8$ tension, though this tension does not impede combining data sets to constrain LiMRs.}
    \label{fig:lensing_degeneracy}
\end{figure*}

Of course, the structure of these degeneracies depend strongly on the types of data involved in the analysis, and we now discuss how the limits we infer are effected by the datasets we employ. For this, we focus on the specific case of a  Weyl LiMR at mininum temperature, and obtain posteriors assuming different combinations of observables, which we show in Fig.~\ref{fig:lensing_degeneracy}.

In the case where we consider only {\it Planck} and BOSS data (i.e., no weak lensing), we see a marked correlation between $\sigma_8$ and $\omega_{\rm cdm}$ in the presence of LiMRs. The degeneracy line, which preserves total matter abundance, extends the LiMR mass $m_X$ contours up to 20 eV, which severely erodes the constraints. This indicates that while CMB and LSS data fix a preferred value of $\omega_m$, it is largely insensitive to the makeup of the matter budget.  Weak-lensing data (from CFHTLens, in this work), in contrast, prefers a specific value of $\omega_{\rm cdm}$ close to the $\Lambda$CDM (no-LiMR) case, which pushes $\sigma_8$ up and $m_X$ closer to zero. This preference for larger $\omega_{\rm cdm}$, at the cost of a higher $\sigma_8$ (which is otherwise disfavored by weak-lensing data~\cite{Heymans:2013fya,DES:2021wwk,Hildebrandt:2018yau}), is what allows us to employ these data to strongly constrain LiMRs.

\begin{figure*}
    \centering
    \includegraphics[width=\linewidth]{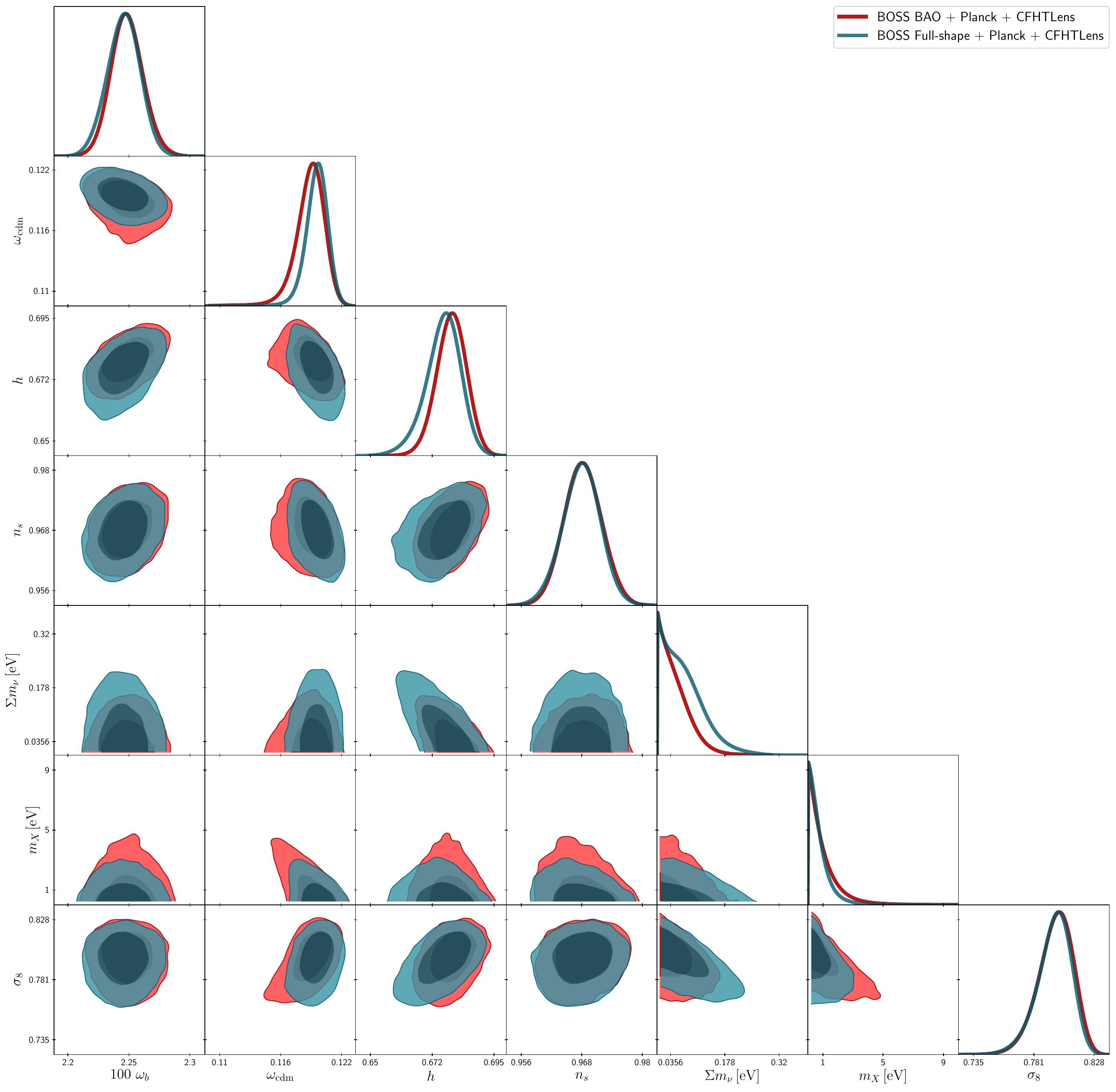}
    \caption{ Corner plot of 1- and 2-D posteriors of all parameters governing a cosmology with a minimum-temperature Weyl relic, for analyses including BAO-only LSS information ({\tt P18 + BOSS-BAO + WLens}, red) and including broadband LSS information ({\tt P18 + BOSS-FS +WLens}, blue). A more stringent measurement on clustering matter by the latter dataset disfavors excess displacement of CDM abundance by light relics, giving a stronger constraint on $m_X$. Also contributing is a slight preference for smaller $h$ by the full-shape dataset, which influences the relic-mass constraint via degeneracies with neutrino mass.
    }
    \label{fig:FS_vs_BAO}
\end{figure*}

With the inclusion of weak-lensing data to break the bulk of the $\omega_{\rm cdm} - \omega_X$ degeneracy, the upgrade from BAO to broadband LSS data further improves our constraining power, strengthening our bounds on $m_X$ by $\sim 30\%$. Fig.~\ref{fig:FS_vs_BAO} displays 2D contours that compare the two analyses. This improvement mainly stems from a better measurement of clustering (CDM + baryon) matter in the universe, more strongly disfavoring the displacement of CDM for LiMRs. 
An additional refinement follows from the preference of the {\tt P18 + BOSS-FS} data for a slightly smaller $H_0$ compared to the {\tt P18 + BOSS-BAO} combination, an observation also documented in Ref.~\cite{Philcox:2020vvt}. This shift in preference in conjunction with the $h-\sum m_\nu$ degeneracy then allows the {\tt FS} data to accommodate larger neutrino masses, and the modest $m_X - \sum m_\nu$ anti-correlation then pulls for overall smaller relic masses.  It is thus the aggregate combination of CMB, broadband LSS, and weak-lensing data that fully empowers us to obtain the strongest constraints to date on these massive light relics.

A final, technical caveat is that our minimum-temperature constraints assume the existence of a relic, even if massless.
This is because even the $m_X=0$ limit (for fixed $T_X^{(0)}=0.91$ K) corresponds to a fixed amount of radiation energy ($N_{\rm eff}$) injected into a $\Lambda$CDM background, and thus the no-relic $\Lambda$CDM case is not represented within this parameter space. 
This would slightly shift our constraints, 
but as the amounts of $N_{\rm eff}$ contributed by the minimum-temperature case ($\Delta N_{\rm eff}<0.1$ even in the Dirac scenario) cannot be resolved by current data, we expect that these shifts will be small. For the next generation of CMB experiments that are expected to measure $N_{\rm eff}$ contributions at the 0.03 level~\cite{Abazajian:2019eic}, however, this may be a significant consideration.




 \bibliography{LiMRs}

\begin{thebibliography}{71}%
\makeatletter
\providecommand \@ifxundefined [1]{%
 \@ifx{#1\undefined}
}%
\providecommand \@ifnum [1]{%
 \ifnum #1\expandafter \@firstoftwo
 \else \expandafter \@secondoftwo
 \fi
}%
\providecommand \@ifx [1]{%
 \ifx #1\expandafter \@firstoftwo
 \else \expandafter \@secondoftwo
 \fi
}%
\providecommand \natexlab [1]{#1}%
\providecommand \enquote  [1]{``#1''}%
\providecommand \bibnamefont  [1]{#1}%
\providecommand \bibfnamefont [1]{#1}%
\providecommand \citenamefont [1]{#1}%
\providecommand \href@noop [0]{\@secondoftwo}%
\providecommand \href [0]{\begingroup \@sanitize@url \@href}%
\providecommand \@href[1]{\@@startlink{#1}\@@href}%
\providecommand \@@href[1]{\endgroup#1\@@endlink}%
\providecommand \@sanitize@url [0]{\catcode `\\12\catcode `\$12\catcode
  `\&12\catcode `\#12\catcode `\^12\catcode `\_12\catcode `\%12\relax}%
\providecommand \@@startlink[1]{}%
\providecommand \@@endlink[0]{}%
\providecommand \url  [0]{\begingroup\@sanitize@url \@url }%
\providecommand \@url [1]{\endgroup\@href {#1}{\urlprefix }}%
\providecommand \urlprefix  [0]{URL }%
\providecommand \Eprint [0]{\href }%
\providecommand \doibase [0]{https://doi.org/}%
\providecommand \selectlanguage [0]{\@gobble}%
\providecommand \bibinfo  [0]{\@secondoftwo}%
\providecommand \bibfield  [0]{\@secondoftwo}%
\providecommand \translation [1]{[#1]}%
\providecommand \BibitemOpen [0]{}%
\providecommand \bibitemStop [0]{}%
\providecommand \bibitemNoStop [0]{.\EOS\space}%
\providecommand \EOS [0]{\spacefactor3000\relax}%
\providecommand \BibitemShut  [1]{\csname bibitem#1\endcsname}%
\let\auto@bib@innerbib\@empty
\bibitem [{\citenamefont {Aghanim}\ \emph
  {et~al.}(2020{\natexlab{a}})\citenamefont {Aghanim} \emph
  {et~al.}}]{Aghanim:2018eyx}%
  \BibitemOpen
  \bibfield  {author} {\bibinfo {author} {\bibfnamefont {N.}~\bibnamefont
  {Aghanim}} \emph {et~al.} (\bibinfo {collaboration} {Planck}),\ }\bibfield
  {title} {\bibinfo {title} {{Planck 2018 results. VI. Cosmological
  parameters}},\ }\href {https://doi.org/10.1051/0004-6361/201833910}
  {\bibfield  {journal} {\bibinfo  {journal} {Astron. Astrophys.}\ }\textbf
  {\bibinfo {volume} {641}},\ \bibinfo {pages} {A6} (\bibinfo {year}
  {2020}{\natexlab{a}})},\ \Eprint {https://arxiv.org/abs/1807.06209}
  {arXiv:1807.06209 [astro-ph.CO]} \BibitemShut {NoStop}%
\bibitem [{\citenamefont {Alam}\ \emph {et~al.}(2017)\citenamefont {Alam} \emph
  {et~al.}}]{Alam:2016hwk}%
  \BibitemOpen
  \bibfield  {author} {\bibinfo {author} {\bibfnamefont {S.}~\bibnamefont
  {Alam}} \emph {et~al.} (\bibinfo {collaboration} {BOSS}),\ }\bibfield
  {title} {\bibinfo {title} {{The clustering of galaxies in the completed
  SDSS-III Baryon Oscillation Spectroscopic Survey: cosmological analysis of
  the DR12 galaxy sample}},\ }\href {https://doi.org/10.1093/mnras/stx721}
  {\bibfield  {journal} {\bibinfo  {journal} {Mon. Not. Roy. Astron. Soc.}\
  }\textbf {\bibinfo {volume} {470}},\ \bibinfo {pages} {2617} (\bibinfo {year}
  {2017})},\ \Eprint {https://arxiv.org/abs/1607.03155} {arXiv:1607.03155
  [astro-ph.CO]} \BibitemShut {NoStop}%
\bibitem [{\citenamefont {Erben}\ \emph {et~al.}(2013)\citenamefont {Erben}
  \emph {et~al.}}]{Erben:2012zw}%
  \BibitemOpen
  \bibfield  {author} {\bibinfo {author} {\bibfnamefont {T.}~\bibnamefont
  {Erben}} \emph {et~al.},\ }\bibfield  {title} {\bibinfo {title} {{CFHTLenS:
  The Canada-France-Hawaii Telescope Lensing Survey - Imaging Data and
  Catalogue Products}},\ }\href {https://doi.org/10.1093/mnras/stt928}
  {\bibfield  {journal} {\bibinfo  {journal} {Mon. Not. Roy. Astron. Soc.}\
  }\textbf {\bibinfo {volume} {433}},\ \bibinfo {pages} {2545} (\bibinfo {year}
  {2013})},\ \Eprint {https://arxiv.org/abs/1210.8156} {arXiv:1210.8156
  [astro-ph.CO]} \BibitemShut {NoStop}%
\bibitem [{\citenamefont {Spergel}\ and\ \citenamefont
  {Steinhardt}(2000)}]{Spergel:1999mh}%
  \BibitemOpen
  \bibfield  {author} {\bibinfo {author} {\bibfnamefont {D.~N.}\ \bibnamefont
  {Spergel}}\ and\ \bibinfo {author} {\bibfnamefont {P.~J.}\ \bibnamefont
  {Steinhardt}},\ }\bibfield  {title} {\bibinfo {title} {{Observational
  evidence for selfinteracting cold dark matter}},\ }\href
  {https://doi.org/10.1103/PhysRevLett.84.3760} {\bibfield  {journal} {\bibinfo
   {journal} {Phys. Rev. Lett.}\ }\textbf {\bibinfo {volume} {84}},\ \bibinfo
  {pages} {3760} (\bibinfo {year} {2000})},\ \Eprint
  {https://arxiv.org/abs/astro-ph/9909386} {arXiv:astro-ph/9909386 [astro-ph]}
  \BibitemShut {NoStop}%
\bibitem [{\citenamefont {Buen-Abad}\ \emph {et~al.}(2015)\citenamefont
  {Buen-Abad}, \citenamefont {Marques-Tavares},\ and\ \citenamefont
  {Schmaltz}}]{Buen-Abad:2015ova}%
  \BibitemOpen
  \bibfield  {author} {\bibinfo {author} {\bibfnamefont {M.~A.}\ \bibnamefont
  {Buen-Abad}}, \bibinfo {author} {\bibfnamefont {G.}~\bibnamefont
  {Marques-Tavares}},\ and\ \bibinfo {author} {\bibfnamefont {M.}~\bibnamefont
  {Schmaltz}},\ }\bibfield  {title} {\bibinfo {title} {{Non-Abelian dark matter
  and dark radiation}},\ }\href {https://doi.org/10.1103/PhysRevD.92.023531}
  {\bibfield  {journal} {\bibinfo  {journal} {Phys. Rev. D}\ }\textbf {\bibinfo
  {volume} {92}},\ \bibinfo {pages} {023531} (\bibinfo {year} {2015})},\
  \Eprint {https://arxiv.org/abs/1505.03542} {arXiv:1505.03542 [hep-ph]}
  \BibitemShut {NoStop}%
\bibitem [{\citenamefont {Kamada}\ \emph {et~al.}(2017)\citenamefont {Kamada},
  \citenamefont {Kaplinghat}, \citenamefont {Pace},\ and\ \citenamefont
  {Yu}}]{Kamada:2016euw}%
  \BibitemOpen
  \bibfield  {author} {\bibinfo {author} {\bibfnamefont {A.}~\bibnamefont
  {Kamada}}, \bibinfo {author} {\bibfnamefont {M.}~\bibnamefont {Kaplinghat}},
  \bibinfo {author} {\bibfnamefont {A.~B.}\ \bibnamefont {Pace}},\ and\
  \bibinfo {author} {\bibfnamefont {H.-B.}\ \bibnamefont {Yu}},\ }\bibfield
  {title} {\bibinfo {title} {{How the Self-Interacting Dark Matter Model
  Explains the Diverse Galactic Rotation Curves}},\ }\href
  {https://doi.org/10.1103/PhysRevLett.119.111102} {\bibfield  {journal}
  {\bibinfo  {journal} {Phys. Rev. Lett.}\ }\textbf {\bibinfo {volume} {119}},\
  \bibinfo {pages} {111102} (\bibinfo {year} {2017})},\ \Eprint
  {https://arxiv.org/abs/1611.02716} {arXiv:1611.02716 [astro-ph.GA]}
  \BibitemShut {NoStop}%
\bibitem [{\citenamefont {Park}\ \emph {et~al.}(2019)\citenamefont {Park},
  \citenamefont {Kreisch}, \citenamefont {Dunkley}, \citenamefont
  {Hadzhiyska},\ and\ \citenamefont {Cyr-Racine}}]{Park:2019ibn}%
  \BibitemOpen
  \bibfield  {author} {\bibinfo {author} {\bibfnamefont {M.}~\bibnamefont
  {Park}}, \bibinfo {author} {\bibfnamefont {C.~D.}\ \bibnamefont {Kreisch}},
  \bibinfo {author} {\bibfnamefont {J.}~\bibnamefont {Dunkley}}, \bibinfo
  {author} {\bibfnamefont {B.}~\bibnamefont {Hadzhiyska}},\ and\ \bibinfo
  {author} {\bibfnamefont {F.-Y.}\ \bibnamefont {Cyr-Racine}},\ }\bibfield
  {title} {\bibinfo {title} {{$\Lambda$CDM or self-interacting neutrinos: How
  CMB data can tell the two models apart}},\ }\href
  {https://doi.org/10.1103/PhysRevD.100.063524} {\bibfield  {journal} {\bibinfo
   {journal} {Phys. Rev. D}\ }\textbf {\bibinfo {volume} {100}},\ \bibinfo
  {pages} {063524} (\bibinfo {year} {2019})},\ \Eprint
  {https://arxiv.org/abs/1904.02625} {arXiv:1904.02625 [astro-ph.CO]}
  \BibitemShut {NoStop}%
\bibitem [{\citenamefont {Mangano}\ \emph {et~al.}(2005)\citenamefont
  {Mangano}, \citenamefont {Miele}, \citenamefont {Pastor}, \citenamefont
  {Pinto}, \citenamefont {Pisanti},\ and\ \citenamefont
  {Serpico}}]{Mangano:2005cc}%
  \BibitemOpen
  \bibfield  {author} {\bibinfo {author} {\bibfnamefont {G.}~\bibnamefont
  {Mangano}}, \bibinfo {author} {\bibfnamefont {G.}~\bibnamefont {Miele}},
  \bibinfo {author} {\bibfnamefont {S.}~\bibnamefont {Pastor}}, \bibinfo
  {author} {\bibfnamefont {T.}~\bibnamefont {Pinto}}, \bibinfo {author}
  {\bibfnamefont {O.}~\bibnamefont {Pisanti}},\ and\ \bibinfo {author}
  {\bibfnamefont {P.~D.}\ \bibnamefont {Serpico}},\ }\bibfield  {title}
  {\bibinfo {title} {{Relic neutrino decoupling including flavor
  oscillations}},\ }\href {https://doi.org/10.1016/j.nuclphysb.2005.09.041}
  {\bibfield  {journal} {\bibinfo  {journal} {Nucl. Phys. B}\ }\textbf
  {\bibinfo {volume} {729}},\ \bibinfo {pages} {221} (\bibinfo {year}
  {2005})},\ \Eprint {https://arxiv.org/abs/hep-ph/0506164}
  {arXiv:hep-ph/0506164} \BibitemShut {NoStop}%
\bibitem [{\citenamefont {de~Salas}\ and\ \citenamefont
  {Pastor}(2016)}]{deSalas:2016ztq}%
  \BibitemOpen
  \bibfield  {author} {\bibinfo {author} {\bibfnamefont {P.~F.}\ \bibnamefont
  {de~Salas}}\ and\ \bibinfo {author} {\bibfnamefont {S.}~\bibnamefont
  {Pastor}},\ }\bibfield  {title} {\bibinfo {title} {{Relic neutrino decoupling
  with flavour oscillations revisited}},\ }\href
  {https://doi.org/10.1088/1475-7516/2016/07/051} {\bibfield  {journal}
  {\bibinfo  {journal} {JCAP}\ }\textbf {\bibinfo {volume} {07}},\ \bibinfo
  {pages} {051}},\ \Eprint {https://arxiv.org/abs/1606.06986} {arXiv:1606.06986
  [hep-ph]} \BibitemShut {NoStop}%
\bibitem [{\citenamefont {Weinberg}(1982)}]{Weinberg:1982zq}%
  \BibitemOpen
  \bibfield  {author} {\bibinfo {author} {\bibfnamefont {S.}~\bibnamefont
  {Weinberg}},\ }\bibfield  {title} {\bibinfo {title} {{Cosmological
  Constraints on the Scale of Supersymmetry Breaking}},\ }\href
  {https://doi.org/10.1103/PhysRevLett.48.1303} {\bibfield  {journal} {\bibinfo
   {journal} {Phys. Rev. Lett.}\ }\textbf {\bibinfo {volume} {48}},\ \bibinfo
  {pages} {1303} (\bibinfo {year} {1982})}\BibitemShut {NoStop}%
\bibitem [{\citenamefont {Giudice}\ \emph {et~al.}(1999)\citenamefont
  {Giudice}, \citenamefont {Riotto},\ and\ \citenamefont
  {Tkachev}}]{Giudice:1999am}%
  \BibitemOpen
  \bibfield  {author} {\bibinfo {author} {\bibfnamefont {G.~F.}\ \bibnamefont
  {Giudice}}, \bibinfo {author} {\bibfnamefont {A.}~\bibnamefont {Riotto}},\
  and\ \bibinfo {author} {\bibfnamefont {I.}~\bibnamefont {Tkachev}},\
  }\bibfield  {title} {\bibinfo {title} {{Thermal and nonthermal production of
  gravitinos in the early universe}},\ }\href
  {https://doi.org/10.1088/1126-6708/1999/11/036} {\bibfield  {journal}
  {\bibinfo  {journal} {JHEP}\ }\textbf {\bibinfo {volume} {11}},\ \bibinfo
  {pages} {036}},\ \Eprint {https://arxiv.org/abs/hep-ph/9911302}
  {arXiv:hep-ph/9911302} \BibitemShut {NoStop}%
\bibitem [{\citenamefont {Feng}\ \emph {et~al.}(2010)\citenamefont {Feng},
  \citenamefont {Kamionkowski},\ and\ \citenamefont {Lee}}]{Feng:2010ij}%
  \BibitemOpen
  \bibfield  {author} {\bibinfo {author} {\bibfnamefont {J.~L.}\ \bibnamefont
  {Feng}}, \bibinfo {author} {\bibfnamefont {M.}~\bibnamefont {Kamionkowski}},\
  and\ \bibinfo {author} {\bibfnamefont {S.~K.}\ \bibnamefont {Lee}},\
  }\bibfield  {title} {\bibinfo {title} {{Light Gravitinos at Colliders and
  Implications for Cosmology}},\ }\href
  {https://doi.org/10.1103/PhysRevD.82.015012} {\bibfield  {journal} {\bibinfo
  {journal} {Phys. Rev. D}\ }\textbf {\bibinfo {volume} {82}},\ \bibinfo
  {pages} {015012} (\bibinfo {year} {2010})},\ \Eprint
  {https://arxiv.org/abs/1004.4213} {arXiv:1004.4213 [hep-ph]} \BibitemShut
  {NoStop}%
\bibitem [{\citenamefont {Hook}\ and\ \citenamefont
  {Murayama}(2015)}]{Hook:2015tra}%
  \BibitemOpen
  \bibfield  {author} {\bibinfo {author} {\bibfnamefont {A.}~\bibnamefont
  {Hook}}\ and\ \bibinfo {author} {\bibfnamefont {H.}~\bibnamefont
  {Murayama}},\ }\bibfield  {title} {\bibinfo {title} {{Low-energy
  Supersymmetry Breaking Without the Gravitino Problem}},\ }\href
  {https://doi.org/10.1103/PhysRevD.92.015004} {\bibfield  {journal} {\bibinfo
  {journal} {Phys. Rev. D}\ }\textbf {\bibinfo {volume} {92}},\ \bibinfo
  {pages} {015004} (\bibinfo {year} {2015})},\ \Eprint
  {https://arxiv.org/abs/1503.04880} {arXiv:1503.04880 [hep-ph]} \BibitemShut
  {NoStop}%
\bibitem [{\citenamefont {Hook}\ \emph {et~al.}(2018)\citenamefont {Hook},
  \citenamefont {McGehee},\ and\ \citenamefont {Murayama}}]{Hook:2018sai}%
  \BibitemOpen
  \bibfield  {author} {\bibinfo {author} {\bibfnamefont {A.}~\bibnamefont
  {Hook}}, \bibinfo {author} {\bibfnamefont {R.}~\bibnamefont {McGehee}},\ and\
  \bibinfo {author} {\bibfnamefont {H.}~\bibnamefont {Murayama}},\ }\bibfield
  {title} {\bibinfo {title} {{Cosmologically Viable Low-energy Supersymmetry
  Breaking}},\ }\href {https://doi.org/10.1103/PhysRevD.98.115036} {\bibfield
  {journal} {\bibinfo  {journal} {Phys. Rev. D}\ }\textbf {\bibinfo {volume}
  {98}},\ \bibinfo {pages} {115036} (\bibinfo {year} {2018})},\ \Eprint
  {https://arxiv.org/abs/1801.10160} {arXiv:1801.10160 [hep-ph]} \BibitemShut
  {NoStop}%
\bibitem [{\citenamefont {Ackerman}\ \emph {et~al.}(2009)\citenamefont
  {Ackerman}, \citenamefont {Buckley}, \citenamefont {Carroll},\ and\
  \citenamefont {Kamionkowski}}]{Ackerman:mha}%
  \BibitemOpen
  \bibfield  {author} {\bibinfo {author} {\bibfnamefont {L.}~\bibnamefont
  {Ackerman}}, \bibinfo {author} {\bibfnamefont {M.~R.}\ \bibnamefont
  {Buckley}}, \bibinfo {author} {\bibfnamefont {S.~M.}\ \bibnamefont
  {Carroll}},\ and\ \bibinfo {author} {\bibfnamefont {M.}~\bibnamefont
  {Kamionkowski}},\ }\bibfield  {title} {\bibinfo {title} {{Dark Matter and
  Dark Radiation}},\ }\href {https://doi.org/10.1103/PhysRevD.79.023519}
  {\bibfield  {journal} {\bibinfo  {journal} {Phys. Rev. D}\ }\textbf {\bibinfo
  {volume} {79}},\ \bibinfo {pages} {023519} (\bibinfo {year} {2009})},\
  \Eprint {https://arxiv.org/abs/0810.5126} {arXiv:0810.5126 [hep-ph]}
  \BibitemShut {NoStop}%
\bibitem [{\citenamefont {Vogel}\ and\ \citenamefont
  {Redondo}(2014)}]{Vogel:2013raa}%
  \BibitemOpen
  \bibfield  {author} {\bibinfo {author} {\bibfnamefont {H.}~\bibnamefont
  {Vogel}}\ and\ \bibinfo {author} {\bibfnamefont {J.}~\bibnamefont
  {Redondo}},\ }\bibfield  {title} {\bibinfo {title} {{Dark Radiation
  constraints on minicharged particles in models with a hidden photon}},\
  }\href {https://doi.org/10.1088/1475-7516/2014/02/029} {\bibfield  {journal}
  {\bibinfo  {journal} {JCAP}\ }\textbf {\bibinfo {volume} {02}},\ \bibinfo
  {pages} {029}},\ \Eprint {https://arxiv.org/abs/1311.2600} {arXiv:1311.2600
  [hep-ph]} \BibitemShut {NoStop}%
\bibitem [{\citenamefont {Peccei}\ and\ \citenamefont
  {Quinn}(1977)}]{Peccei:1977hh}%
  \BibitemOpen
  \bibfield  {author} {\bibinfo {author} {\bibfnamefont {R.~D.}\ \bibnamefont
  {Peccei}}\ and\ \bibinfo {author} {\bibfnamefont {H.~R.}\ \bibnamefont
  {Quinn}},\ }\bibfield  {title} {\bibinfo {title} {{CP Conservation in the
  Presence of Instantons}},\ }\href
  {https://doi.org/10.1103/PhysRevLett.38.1440} {\bibfield  {journal} {\bibinfo
   {journal} {Phys. Rev. Lett.}\ }\textbf {\bibinfo {volume} {38}},\ \bibinfo
  {pages} {1440} (\bibinfo {year} {1977})}\BibitemShut {NoStop}%
\bibitem [{\citenamefont {Arvanitaki}\ \emph {et~al.}(2010)\citenamefont
  {Arvanitaki}, \citenamefont {Dimopoulos}, \citenamefont {Dubovsky},
  \citenamefont {Kaloper},\ and\ \citenamefont
  {March-Russell}}]{Arvanitaki:2009fg}%
  \BibitemOpen
  \bibfield  {author} {\bibinfo {author} {\bibfnamefont {A.}~\bibnamefont
  {Arvanitaki}}, \bibinfo {author} {\bibfnamefont {S.}~\bibnamefont
  {Dimopoulos}}, \bibinfo {author} {\bibfnamefont {S.}~\bibnamefont
  {Dubovsky}}, \bibinfo {author} {\bibfnamefont {N.}~\bibnamefont {Kaloper}},\
  and\ \bibinfo {author} {\bibfnamefont {J.}~\bibnamefont {March-Russell}},\
  }\bibfield  {title} {\bibinfo {title} {{String Axiverse}},\ }\href
  {https://doi.org/10.1103/PhysRevD.81.123530} {\bibfield  {journal} {\bibinfo
  {journal} {Phys. Rev. D}\ }\textbf {\bibinfo {volume} {81}},\ \bibinfo
  {pages} {123530} (\bibinfo {year} {2010})},\ \Eprint
  {https://arxiv.org/abs/0905.4720} {arXiv:0905.4720 [hep-th]} \BibitemShut
  {NoStop}%
\bibitem [{\citenamefont {Marsh}(2016)}]{Marsh:2015xka}%
  \BibitemOpen
  \bibfield  {author} {\bibinfo {author} {\bibfnamefont {D.~J.~E.}\
  \bibnamefont {Marsh}},\ }\bibfield  {title} {\bibinfo {title} {{Axion
  Cosmology}},\ }\href {https://doi.org/10.1016/j.physrep.2016.06.005}
  {\bibfield  {journal} {\bibinfo  {journal} {Phys. Rept.}\ }\textbf {\bibinfo
  {volume} {643}},\ \bibinfo {pages} {1} (\bibinfo {year} {2016})},\ \Eprint
  {https://arxiv.org/abs/1510.07633} {arXiv:1510.07633 [astro-ph.CO]}
  \BibitemShut {NoStop}%
\bibitem [{\citenamefont {Baumann}\ \emph {et~al.}(2016)\citenamefont
  {Baumann}, \citenamefont {Green},\ and\ \citenamefont
  {Wallisch}}]{Baumann:2016wac}%
  \BibitemOpen
  \bibfield  {author} {\bibinfo {author} {\bibfnamefont {D.}~\bibnamefont
  {Baumann}}, \bibinfo {author} {\bibfnamefont {D.}~\bibnamefont {Green}},\
  and\ \bibinfo {author} {\bibfnamefont {B.}~\bibnamefont {Wallisch}},\
  }\bibfield  {title} {\bibinfo {title} {{New Target for Cosmic Axion
  Searches}},\ }\href {https://doi.org/10.1103/PhysRevLett.117.171301}
  {\bibfield  {journal} {\bibinfo  {journal} {Phys. Rev. Lett.}\ }\textbf
  {\bibinfo {volume} {117}},\ \bibinfo {pages} {171301} (\bibinfo {year}
  {2016})},\ \Eprint {https://arxiv.org/abs/1604.08614} {arXiv:1604.08614
  [astro-ph.CO]} \BibitemShut {NoStop}%
\bibitem [{\citenamefont {Boyarsky}\ \emph
  {et~al.}(2009{\natexlab{a}})\citenamefont {Boyarsky}, \citenamefont
  {Ruchayskiy},\ and\ \citenamefont {Shaposhnikov}}]{Boyarsky:2009ix}%
  \BibitemOpen
  \bibfield  {author} {\bibinfo {author} {\bibfnamefont {A.}~\bibnamefont
  {Boyarsky}}, \bibinfo {author} {\bibfnamefont {O.}~\bibnamefont
  {Ruchayskiy}},\ and\ \bibinfo {author} {\bibfnamefont {M.}~\bibnamefont
  {Shaposhnikov}},\ }\bibfield  {title} {\bibinfo {title} {{The Role of sterile
  neutrinos in cosmology and astrophysics}},\ }\href
  {https://doi.org/10.1146/annurev.nucl.010909.083654} {\bibfield  {journal}
  {\bibinfo  {journal} {Ann. Rev. Nucl. Part. Sci.}\ }\textbf {\bibinfo
  {volume} {59}},\ \bibinfo {pages} {191} (\bibinfo {year}
  {2009}{\natexlab{a}})},\ \Eprint {https://arxiv.org/abs/0901.0011}
  {arXiv:0901.0011 [hep-ph]} \BibitemShut {NoStop}%
\bibitem [{\citenamefont {Abazajian}\ \emph {et~al.}(2012)\citenamefont
  {Abazajian} \emph {et~al.}}]{Abazajian:2012ys}%
  \BibitemOpen
  \bibfield  {author} {\bibinfo {author} {\bibfnamefont {K.~N.}\ \bibnamefont
  {Abazajian}} \emph {et~al.},\ }\href@noop {} {\bibinfo {title} {{Light
  Sterile Neutrinos: A White Paper}}} (\bibinfo {year} {2012}),\ \Eprint
  {https://arxiv.org/abs/1204.5379} {arXiv:1204.5379 [hep-ph]} \BibitemShut
  {NoStop}%
\bibitem [{\citenamefont {Brust}\ \emph {et~al.}(2013)\citenamefont {Brust},
  \citenamefont {Kaplan},\ and\ \citenamefont {Walters}}]{Brust:2013ova}%
  \BibitemOpen
  \bibfield  {author} {\bibinfo {author} {\bibfnamefont {C.}~\bibnamefont
  {Brust}}, \bibinfo {author} {\bibfnamefont {D.~E.}\ \bibnamefont {Kaplan}},\
  and\ \bibinfo {author} {\bibfnamefont {M.~T.}\ \bibnamefont {Walters}},\
  }\bibfield  {title} {\bibinfo {title} {{New Light Species and the CMB}},\
  }\href {https://doi.org/10.1007/JHEP12(2013)058} {\bibfield  {journal}
  {\bibinfo  {journal} {JHEP}\ }\textbf {\bibinfo {volume} {12}},\ \bibinfo
  {pages} {058}},\ \Eprint {https://arxiv.org/abs/1303.5379} {arXiv:1303.5379
  [hep-ph]} \BibitemShut {NoStop}%
\bibitem [{\citenamefont {Baumann}\ \emph {et~al.}(2018)\citenamefont
  {Baumann}, \citenamefont {Green},\ and\ \citenamefont
  {Wallisch}}]{Baumann:2017gkg}%
  \BibitemOpen
  \bibfield  {author} {\bibinfo {author} {\bibfnamefont {D.}~\bibnamefont
  {Baumann}}, \bibinfo {author} {\bibfnamefont {D.}~\bibnamefont {Green}},\
  and\ \bibinfo {author} {\bibfnamefont {B.}~\bibnamefont {Wallisch}},\
  }\bibfield  {title} {\bibinfo {title} {{Searching for light relics with
  large-scale structure}},\ }\href
  {https://doi.org/10.1088/1475-7516/2018/08/029} {\bibfield  {journal}
  {\bibinfo  {journal} {JCAP}\ }\textbf {\bibinfo {volume} {08}},\ \bibinfo
  {pages} {029}},\ \Eprint {https://arxiv.org/abs/1712.08067} {arXiv:1712.08067
  [astro-ph.CO]} \BibitemShut {NoStop}%
\bibitem [{\citenamefont {Green}\ \emph {et~al.}(2019)\citenamefont {Green}
  \emph {et~al.}}]{Green:2019glg}%
  \BibitemOpen
  \bibfield  {author} {\bibinfo {author} {\bibfnamefont {D.}~\bibnamefont
  {Green}} \emph {et~al.},\ }\bibfield  {title} {\bibinfo {title} {{Messengers
  from the Early Universe: Cosmic Neutrinos and Other Light Relics}},\
  }\href@noop {} {\bibfield  {journal} {\bibinfo  {journal} {Bull. Am. Astron.
  Soc.}\ }\textbf {\bibinfo {volume} {51}},\ \bibinfo {pages} {159} (\bibinfo
  {year} {2019})},\ \Eprint {https://arxiv.org/abs/1903.04763}
  {arXiv:1903.04763 [astro-ph.CO]} \BibitemShut {NoStop}%
\bibitem [{\citenamefont {Bell}\ \emph {et~al.}(2006)\citenamefont {Bell},
  \citenamefont {Pierpaoli},\ and\ \citenamefont {Sigurdson}}]{Bell:2005dr}%
  \BibitemOpen
  \bibfield  {author} {\bibinfo {author} {\bibfnamefont {N.~F.}\ \bibnamefont
  {Bell}}, \bibinfo {author} {\bibfnamefont {E.}~\bibnamefont {Pierpaoli}},\
  and\ \bibinfo {author} {\bibfnamefont {K.}~\bibnamefont {Sigurdson}},\
  }\bibfield  {title} {\bibinfo {title} {{Cosmological signatures of
  interacting neutrinos}},\ }\href {https://doi.org/10.1103/PhysRevD.73.063523}
  {\bibfield  {journal} {\bibinfo  {journal} {Phys. Rev. D}\ }\textbf {\bibinfo
  {volume} {73}},\ \bibinfo {pages} {063523} (\bibinfo {year} {2006})},\
  \Eprint {https://arxiv.org/abs/astro-ph/0511410} {arXiv:astro-ph/0511410}
  \BibitemShut {NoStop}%
\bibitem [{\citenamefont {Brinckmann}\ \emph {et~al.}(2020)\citenamefont
  {Brinckmann}, \citenamefont {Chang},\ and\ \citenamefont
  {LoVerde}}]{Brinckmann:2020bcn}%
  \BibitemOpen
  \bibfield  {author} {\bibinfo {author} {\bibfnamefont {T.}~\bibnamefont
  {Brinckmann}}, \bibinfo {author} {\bibfnamefont {J.~H.}\ \bibnamefont
  {Chang}},\ and\ \bibinfo {author} {\bibfnamefont {M.}~\bibnamefont
  {LoVerde}},\ }\href@noop {} {\bibinfo {title} {{Self-interacting neutrinos,
  the Hubble parameter tension, and the Cosmic Microwave Background}}}
  (\bibinfo {year} {2020}),\ \Eprint {https://arxiv.org/abs/2012.11830}
  {arXiv:2012.11830 [astro-ph.CO]} \BibitemShut {NoStop}%
\bibitem [{\citenamefont {Aker}\ \emph {et~al.}(2019)\citenamefont {Aker} \emph
  {et~al.}}]{Aker:2019uuj}%
  \BibitemOpen
  \bibfield  {author} {\bibinfo {author} {\bibfnamefont {M.}~\bibnamefont
  {Aker}} \emph {et~al.} (\bibinfo {collaboration} {KATRIN}),\ }\bibfield
  {title} {\bibinfo {title} {{Improved Upper Limit on the Neutrino Mass from a
  Direct Kinematic Method by KATRIN}},\ }\href
  {https://doi.org/10.1103/PhysRevLett.123.221802} {\bibfield  {journal}
  {\bibinfo  {journal} {Phys. Rev. Lett.}\ }\textbf {\bibinfo {volume} {123}},\
  \bibinfo {pages} {221802} (\bibinfo {year} {2019})},\ \Eprint
  {https://arxiv.org/abs/1909.06048} {arXiv:1909.06048 [hep-ex]} \BibitemShut
  {NoStop}%
\bibitem [{\citenamefont {Boyarsky}\ \emph
  {et~al.}(2009{\natexlab{b}})\citenamefont {Boyarsky}, \citenamefont
  {Lesgourgues}, \citenamefont {Ruchayskiy},\ and\ \citenamefont
  {Viel}}]{Boyarsky:2008xj}%
  \BibitemOpen
  \bibfield  {author} {\bibinfo {author} {\bibfnamefont {A.}~\bibnamefont
  {Boyarsky}}, \bibinfo {author} {\bibfnamefont {J.}~\bibnamefont
  {Lesgourgues}}, \bibinfo {author} {\bibfnamefont {O.}~\bibnamefont
  {Ruchayskiy}},\ and\ \bibinfo {author} {\bibfnamefont {M.}~\bibnamefont
  {Viel}},\ }\bibfield  {title} {\bibinfo {title} {{Lyman-alpha constraints on
  warm and on warm-plus-cold dark matter models}},\ }\href
  {https://doi.org/10.1088/1475-7516/2009/05/012} {\bibfield  {journal}
  {\bibinfo  {journal} {JCAP}\ }\textbf {\bibinfo {volume} {05}},\ \bibinfo
  {pages} {012}},\ \Eprint {https://arxiv.org/abs/0812.0010} {arXiv:0812.0010
  [astro-ph]} \BibitemShut {NoStop}%
\bibitem [{\citenamefont {Mu\~noz}\ and\ \citenamefont
  {Dvorkin}(2018)}]{Munoz:2018ajr}%
  \BibitemOpen
  \bibfield  {author} {\bibinfo {author} {\bibfnamefont {J.~B.}\ \bibnamefont
  {Mu\~noz}}\ and\ \bibinfo {author} {\bibfnamefont {C.}~\bibnamefont
  {Dvorkin}},\ }\bibfield  {title} {\bibinfo {title} {{Efficient Computation of
  Galaxy Bias with Neutrinos and Other Relics}},\ }\href
  {https://doi.org/10.1103/PhysRevD.98.043503} {\bibfield  {journal} {\bibinfo
  {journal} {Phys. Rev. D}\ }\textbf {\bibinfo {volume} {98}},\ \bibinfo
  {pages} {043503} (\bibinfo {year} {2018})},\ \Eprint
  {https://arxiv.org/abs/1805.11623} {arXiv:1805.11623 [astro-ph.CO]}
  \BibitemShut {NoStop}%
\bibitem [{\citenamefont {DePorzio}\ \emph {et~al.}(2021)\citenamefont
  {DePorzio}, \citenamefont {Xu}, \citenamefont {Mu\~noz},\ and\ \citenamefont
  {Dvorkin}}]{DePorzio:2020wcz}%
  \BibitemOpen
  \bibfield  {author} {\bibinfo {author} {\bibfnamefont {N.}~\bibnamefont
  {DePorzio}}, \bibinfo {author} {\bibfnamefont {W.~L.}\ \bibnamefont {Xu}},
  \bibinfo {author} {\bibfnamefont {J.~B.}\ \bibnamefont {Mu\~noz}},\ and\
  \bibinfo {author} {\bibfnamefont {C.}~\bibnamefont {Dvorkin}},\ }\bibfield
  {title} {\bibinfo {title} {{Finding eV-scale light relics with cosmological
  observables}},\ }\href {https://doi.org/10.1103/PhysRevD.103.023504}
  {\bibfield  {journal} {\bibinfo  {journal} {Phys. Rev. D}\ }\textbf {\bibinfo
  {volume} {103}},\ \bibinfo {pages} {023504} (\bibinfo {year} {2021})},\
  \Eprint {https://arxiv.org/abs/2006.09380} {arXiv:2006.09380 [astro-ph.CO]}
  \BibitemShut {NoStop}%
\bibitem [{\citenamefont {Bashinsky}\ and\ \citenamefont
  {Seljak}(2004)}]{Bashinsky:2003tk}%
  \BibitemOpen
  \bibfield  {author} {\bibinfo {author} {\bibfnamefont {S.}~\bibnamefont
  {Bashinsky}}\ and\ \bibinfo {author} {\bibfnamefont {U.}~\bibnamefont
  {Seljak}},\ }\bibfield  {title} {\bibinfo {title} {{Neutrino perturbations in
  CMB anisotropy and matter clustering}},\ }\href
  {https://doi.org/10.1103/PhysRevD.69.083002} {\bibfield  {journal} {\bibinfo
  {journal} {Phys. Rev. D}\ }\textbf {\bibinfo {volume} {69}},\ \bibinfo
  {pages} {083002} (\bibinfo {year} {2004})},\ \Eprint
  {https://arxiv.org/abs/astro-ph/0310198} {arXiv:astro-ph/0310198}
  \BibitemShut {NoStop}%
\bibitem [{\citenamefont {Ade}\ \emph {et~al.}(2019)\citenamefont {Ade} \emph
  {et~al.}}]{Ade:2018sbj}%
  \BibitemOpen
  \bibfield  {author} {\bibinfo {author} {\bibfnamefont {P.}~\bibnamefont
  {Ade}} \emph {et~al.} (\bibinfo {collaboration} {Simons Observatory}),\
  }\bibfield  {title} {\bibinfo {title} {{The Simons Observatory: Science goals
  and forecasts}},\ }\href {https://doi.org/10.1088/1475-7516/2019/02/056}
  {\bibfield  {journal} {\bibinfo  {journal} {JCAP}\ }\textbf {\bibinfo
  {volume} {02}},\ \bibinfo {pages} {056}},\ \Eprint
  {https://arxiv.org/abs/1808.07445} {arXiv:1808.07445 [astro-ph.CO]}
  \BibitemShut {NoStop}%
\bibitem [{\citenamefont {Abazajian}\ \emph {et~al.}(2019)\citenamefont
  {Abazajian} \emph {et~al.}}]{Abazajian:2019eic}%
  \BibitemOpen
  \bibfield  {author} {\bibinfo {author} {\bibfnamefont {K.}~\bibnamefont
  {Abazajian}} \emph {et~al.},\ }\href@noop {} {\bibinfo {title} {{CMB-S4
  Science Case, Reference Design, and Project Plan}}} (\bibinfo {year}
  {2019}),\ \Eprint {https://arxiv.org/abs/1907.04473} {arXiv:1907.04473
  [astro-ph.IM]} \BibitemShut {NoStop}%
\bibitem [{\citenamefont {Lesgourgues}\ and\ \citenamefont
  {Pastor}(2006)}]{Lesgourgues:2006nd}%
  \BibitemOpen
  \bibfield  {author} {\bibinfo {author} {\bibfnamefont {J.}~\bibnamefont
  {Lesgourgues}}\ and\ \bibinfo {author} {\bibfnamefont {S.}~\bibnamefont
  {Pastor}},\ }\bibfield  {title} {\bibinfo {title} {{Massive neutrinos and
  cosmology}},\ }\href {https://doi.org/10.1016/j.physrep.2006.04.001}
  {\bibfield  {journal} {\bibinfo  {journal} {Phys. Rept.}\ }\textbf {\bibinfo
  {volume} {429}},\ \bibinfo {pages} {307} (\bibinfo {year} {2006})},\ \Eprint
  {https://arxiv.org/abs/astro-ph/0603494} {arXiv:astro-ph/0603494}
  \BibitemShut {NoStop}%
\bibitem [{\citenamefont {LoVerde}\ and\ \citenamefont
  {Zaldarriaga}(2014)}]{LoVerde:2013lta}%
  \BibitemOpen
  \bibfield  {author} {\bibinfo {author} {\bibfnamefont {M.}~\bibnamefont
  {LoVerde}}\ and\ \bibinfo {author} {\bibfnamefont {M.}~\bibnamefont
  {Zaldarriaga}},\ }\bibfield  {title} {\bibinfo {title} {{Neutrino clustering
  around spherical dark matter halos}},\ }\href
  {https://doi.org/10.1103/PhysRevD.89.063502} {\bibfield  {journal} {\bibinfo
  {journal} {Phys. Rev. D}\ }\textbf {\bibinfo {volume} {89}},\ \bibinfo
  {pages} {063502} (\bibinfo {year} {2014})},\ \Eprint
  {https://arxiv.org/abs/1310.6459} {arXiv:1310.6459 [astro-ph.CO]}
  \BibitemShut {NoStop}%
\bibitem [{\citenamefont {Ali-Haimoud}\ and\ \citenamefont
  {Bird}(2012)}]{Ali-Haimoud:2012fzp}%
  \BibitemOpen
  \bibfield  {author} {\bibinfo {author} {\bibfnamefont {Y.}~\bibnamefont
  {Ali-Haimoud}}\ and\ \bibinfo {author} {\bibfnamefont {S.}~\bibnamefont
  {Bird}},\ }\bibfield  {title} {\bibinfo {title} {{An efficient implementation
  of massive neutrinos in non-linear structure formation simulations}},\ }\href
  {https://doi.org/10.1093/mnras/sts286} {\bibfield  {journal} {\bibinfo
  {journal} {Mon. Not. Roy. Astron. Soc.}\ }\textbf {\bibinfo {volume} {428}},\
  \bibinfo {pages} {3375} (\bibinfo {year} {2012})},\ \Eprint
  {https://arxiv.org/abs/1209.0461} {arXiv:1209.0461 [astro-ph.CO]}
  \BibitemShut {NoStop}%
\bibitem [{\citenamefont {Chudaykin}\ \emph {et~al.}(2020)\citenamefont
  {Chudaykin}, \citenamefont {Ivanov}, \citenamefont {Philcox},\ and\
  \citenamefont {Simonovi\'c}}]{Chudaykin:2020aoj}%
  \BibitemOpen
  \bibfield  {author} {\bibinfo {author} {\bibfnamefont {A.}~\bibnamefont
  {Chudaykin}}, \bibinfo {author} {\bibfnamefont {M.~M.}\ \bibnamefont
  {Ivanov}}, \bibinfo {author} {\bibfnamefont {O.~H.~E.}\ \bibnamefont
  {Philcox}},\ and\ \bibinfo {author} {\bibfnamefont {M.}~\bibnamefont
  {Simonovi\'c}},\ }\bibfield  {title} {\bibinfo {title} {{Nonlinear
  perturbation theory extension of the Boltzmann code CLASS}},\ }\href
  {https://doi.org/10.1103/PhysRevD.102.063533} {\bibfield  {journal} {\bibinfo
   {journal} {Phys. Rev. D}\ }\textbf {\bibinfo {volume} {102}},\ \bibinfo
  {pages} {063533} (\bibinfo {year} {2020})},\ \Eprint
  {https://arxiv.org/abs/2004.10607} {arXiv:2004.10607 [astro-ph.CO]}
  \BibitemShut {NoStop}%
\bibitem [{\citenamefont {{Blas}}\ \emph {et~al.}(2011)\citenamefont {{Blas}},
  \citenamefont {{Lesgourgues}},\ and\ \citenamefont
  {{Tram}}}]{2011JCAP...07..034B}%
  \BibitemOpen
  \bibfield  {author} {\bibinfo {author} {\bibfnamefont {D.}~\bibnamefont
  {{Blas}}}, \bibinfo {author} {\bibfnamefont {J.}~\bibnamefont
  {{Lesgourgues}}},\ and\ \bibinfo {author} {\bibfnamefont {T.}~\bibnamefont
  {{Tram}}},\ }\href {https://doi.org/10.1088/1475-7516/2011/07/034} {\bibinfo
  {title} {{The Cosmic Linear Anisotropy Solving System (CLASS). Part II:
  Approximation schemes}}} (\bibinfo {year} {2011}),\ \Eprint
  {https://arxiv.org/abs/1104.2933} {arXiv:1104.2933 [astro-ph.CO]}
  \BibitemShut {NoStop}%
\bibitem [{\citenamefont {Villaescusa-Navarro}\ \emph
  {et~al.}(2014)\citenamefont {Villaescusa-Navarro}, \citenamefont {Marulli},
  \citenamefont {Viel}, \citenamefont {Branchini}, \citenamefont {Castorina},
  \citenamefont {Sefusatti},\ and\ \citenamefont
  {Saito}}]{Villaescusa-Navarro:2013pva}%
  \BibitemOpen
  \bibfield  {author} {\bibinfo {author} {\bibfnamefont {F.}~\bibnamefont
  {Villaescusa-Navarro}}, \bibinfo {author} {\bibfnamefont {F.}~\bibnamefont
  {Marulli}}, \bibinfo {author} {\bibfnamefont {M.}~\bibnamefont {Viel}},
  \bibinfo {author} {\bibfnamefont {E.}~\bibnamefont {Branchini}}, \bibinfo
  {author} {\bibfnamefont {E.}~\bibnamefont {Castorina}}, \bibinfo {author}
  {\bibfnamefont {E.}~\bibnamefont {Sefusatti}},\ and\ \bibinfo {author}
  {\bibfnamefont {S.}~\bibnamefont {Saito}},\ }\bibfield  {title} {\bibinfo
  {title} {{Cosmology with massive neutrinos I: towards a realistic modeling of
  the relation between matter, haloes and galaxies}},\ }\href
  {https://doi.org/10.1088/1475-7516/2014/03/011} {\bibfield  {journal}
  {\bibinfo  {journal} {JCAP}\ }\textbf {\bibinfo {volume} {03}},\ \bibinfo
  {pages} {011}},\ \Eprint {https://arxiv.org/abs/1311.0866} {arXiv:1311.0866
  [astro-ph.CO]} \BibitemShut {NoStop}%
\bibitem [{\citenamefont {Biagetti}\ \emph {et~al.}(2014)\citenamefont
  {Biagetti}, \citenamefont {Desjacques}, \citenamefont {Kehagias},\ and\
  \citenamefont {Riotto}}]{Biagetti:2014pha}%
  \BibitemOpen
  \bibfield  {author} {\bibinfo {author} {\bibfnamefont {M.}~\bibnamefont
  {Biagetti}}, \bibinfo {author} {\bibfnamefont {V.}~\bibnamefont
  {Desjacques}}, \bibinfo {author} {\bibfnamefont {A.}~\bibnamefont
  {Kehagias}},\ and\ \bibinfo {author} {\bibfnamefont {A.}~\bibnamefont
  {Riotto}},\ }\bibfield  {title} {\bibinfo {title} {{Nonlocal halo bias with
  and without massive neutrinos}},\ }\href
  {https://doi.org/10.1103/PhysRevD.90.045022} {\bibfield  {journal} {\bibinfo
  {journal} {Phys. Rev. D}\ }\textbf {\bibinfo {volume} {90}},\ \bibinfo
  {pages} {045022} (\bibinfo {year} {2014})},\ \Eprint
  {https://arxiv.org/abs/1405.1435} {arXiv:1405.1435 [astro-ph.CO]}
  \BibitemShut {NoStop}%
\bibitem [{\citenamefont {LoVerde}(2014)}]{LoVerde:2014pxa}%
  \BibitemOpen
  \bibfield  {author} {\bibinfo {author} {\bibfnamefont {M.}~\bibnamefont
  {LoVerde}},\ }\bibfield  {title} {\bibinfo {title} {{Halo bias in mixed dark
  matter cosmologies}},\ }\href {https://doi.org/10.1103/PhysRevD.90.083530}
  {\bibfield  {journal} {\bibinfo  {journal} {Phys. Rev. D}\ }\textbf {\bibinfo
  {volume} {90}},\ \bibinfo {pages} {083530} (\bibinfo {year} {2014})},\
  \Eprint {https://arxiv.org/abs/1405.4855} {arXiv:1405.4855 [astro-ph.CO]}
  \BibitemShut {NoStop}%
\bibitem [{\citenamefont {Ivanov}\ \emph
  {et~al.}(2020{\natexlab{a}})\citenamefont {Ivanov}, \citenamefont
  {Simonovi\'c},\ and\ \citenamefont {Zaldarriaga}}]{Ivanov:2019pdj}%
  \BibitemOpen
  \bibfield  {author} {\bibinfo {author} {\bibfnamefont {M.~M.}\ \bibnamefont
  {Ivanov}}, \bibinfo {author} {\bibfnamefont {M.}~\bibnamefont
  {Simonovi\'c}},\ and\ \bibinfo {author} {\bibfnamefont {M.}~\bibnamefont
  {Zaldarriaga}},\ }\bibfield  {title} {\bibinfo {title} {{Cosmological
  Parameters from the BOSS Galaxy Power Spectrum}},\ }\href
  {https://doi.org/10.1088/1475-7516/2020/05/042} {\bibfield  {journal}
  {\bibinfo  {journal} {JCAP}\ }\textbf {\bibinfo {volume} {05}},\ \bibinfo
  {pages} {042}},\ \Eprint {https://arxiv.org/abs/1909.05277} {arXiv:1909.05277
  [astro-ph.CO]} \BibitemShut {NoStop}%
\bibitem [{\citenamefont {Heymans}\ \emph {et~al.}(2013)\citenamefont {Heymans}
  \emph {et~al.}}]{Heymans:2013fya}%
  \BibitemOpen
  \bibfield  {author} {\bibinfo {author} {\bibfnamefont {C.}~\bibnamefont
  {Heymans}} \emph {et~al.},\ }\bibfield  {title} {\bibinfo {title} {{CFHTLenS
  tomographic weak lensing cosmological parameter constraints: Mitigating the
  impact of intrinsic galaxy alignments}},\ }\href
  {https://doi.org/10.1093/mnras/stt601} {\bibfield  {journal} {\bibinfo
  {journal} {Mon. Not. Roy. Astron. Soc.}\ }\textbf {\bibinfo {volume} {432}},\
  \bibinfo {pages} {2433} (\bibinfo {year} {2013})},\ \Eprint
  {https://arxiv.org/abs/1303.1808} {arXiv:1303.1808 [astro-ph.CO]}
  \BibitemShut {NoStop}%
\bibitem [{\citenamefont {Aghanim}\ \emph
  {et~al.}(2020{\natexlab{b}})\citenamefont {Aghanim} \emph
  {et~al.}}]{Planck:2019nip}%
  \BibitemOpen
  \bibfield  {author} {\bibinfo {author} {\bibfnamefont {N.}~\bibnamefont
  {Aghanim}} \emph {et~al.} (\bibinfo {collaboration} {Planck}),\ }\bibfield
  {title} {\bibinfo {title} {{Planck 2018 results. V. CMB power spectra and
  likelihoods}},\ }\href {https://doi.org/10.1051/0004-6361/201936386}
  {\bibfield  {journal} {\bibinfo  {journal} {Astron. Astrophys.}\ }\textbf
  {\bibinfo {volume} {641}},\ \bibinfo {pages} {A5} (\bibinfo {year}
  {2020}{\natexlab{b}})},\ \Eprint {https://arxiv.org/abs/1907.12875}
  {arXiv:1907.12875 [astro-ph.CO]} \BibitemShut {NoStop}%
\bibitem [{\citenamefont {Philcox}\ \emph {et~al.}(2020)\citenamefont
  {Philcox}, \citenamefont {Ivanov}, \citenamefont {Simonovi\'c},\ and\
  \citenamefont {Zaldarriaga}}]{Philcox:2020vvt}%
  \BibitemOpen
  \bibfield  {author} {\bibinfo {author} {\bibfnamefont {O.~H.~E.}\
  \bibnamefont {Philcox}}, \bibinfo {author} {\bibfnamefont {M.~M.}\
  \bibnamefont {Ivanov}}, \bibinfo {author} {\bibfnamefont {M.}~\bibnamefont
  {Simonovi\'c}},\ and\ \bibinfo {author} {\bibfnamefont {M.}~\bibnamefont
  {Zaldarriaga}},\ }\bibfield  {title} {\bibinfo {title} {{Combining Full-Shape
  and BAO Analyses of Galaxy Power Spectra: A 1.6\textbackslash{}\%
  CMB-independent constraint on H$_0$}},\ }\href
  {https://doi.org/10.1088/1475-7516/2020/05/032} {\bibfield  {journal}
  {\bibinfo  {journal} {JCAP}\ }\textbf {\bibinfo {volume} {05}},\ \bibinfo
  {pages} {032}},\ \Eprint {https://arxiv.org/abs/2002.04035} {arXiv:2002.04035
  [astro-ph.CO]} \BibitemShut {NoStop}%
\bibitem [{\citenamefont {Lemos}\ \emph {et~al.}(2020)\citenamefont {Lemos}
  \emph {et~al.}}]{Lemos:2020jry}%
  \BibitemOpen
  \bibfield  {author} {\bibinfo {author} {\bibfnamefont {P.}~\bibnamefont
  {Lemos}} \emph {et~al.} (\bibinfo {collaboration} {DES}),\ }\href@noop {}
  {\bibinfo {title} {{Assessing tension metrics with Dark Energy Survey and
  Planck data}}} (\bibinfo {year} {2020}),\ \Eprint
  {https://arxiv.org/abs/2012.09554} {arXiv:2012.09554 [astro-ph.CO]}
  \BibitemShut {NoStop}%
\bibitem [{\citenamefont {Abbott}\ \emph
  {et~al.}(2021{\natexlab{a}})\citenamefont {Abbott} \emph
  {et~al.}}]{Abbott:2021bzy}%
  \BibitemOpen
  \bibfield  {author} {\bibinfo {author} {\bibfnamefont {T.~M.~C.}\
  \bibnamefont {Abbott}} \emph {et~al.} (\bibinfo {collaboration} {DES}),\
  }\href@noop {} {\bibinfo {title} {{Dark Energy Survey Year 3 Results:
  Cosmological Constraints from Galaxy Clustering and Weak Lensing}}} (\bibinfo
  {year} {2021}{\natexlab{a}}),\ \Eprint {https://arxiv.org/abs/2105.13549}
  {arXiv:2105.13549 [astro-ph.CO]} \BibitemShut {NoStop}%
\bibitem [{\citenamefont {Viel}\ \emph {et~al.}(2005)\citenamefont {Viel},
  \citenamefont {Lesgourgues}, \citenamefont {Haehnelt}, \citenamefont
  {Matarrese},\ and\ \citenamefont {Riotto}}]{Viel:2005qj}%
  \BibitemOpen
  \bibfield  {author} {\bibinfo {author} {\bibfnamefont {M.}~\bibnamefont
  {Viel}}, \bibinfo {author} {\bibfnamefont {J.}~\bibnamefont {Lesgourgues}},
  \bibinfo {author} {\bibfnamefont {M.~G.}\ \bibnamefont {Haehnelt}}, \bibinfo
  {author} {\bibfnamefont {S.}~\bibnamefont {Matarrese}},\ and\ \bibinfo
  {author} {\bibfnamefont {A.}~\bibnamefont {Riotto}},\ }\bibfield  {title}
  {\bibinfo {title} {{Constraining warm dark matter candidates including
  sterile neutrinos and light gravitinos with WMAP and the Lyman-alpha
  forest}},\ }\href {https://doi.org/10.1103/PhysRevD.71.063534} {\bibfield
  {journal} {\bibinfo  {journal} {Phys. Rev. D}\ }\textbf {\bibinfo {volume}
  {71}},\ \bibinfo {pages} {063534} (\bibinfo {year} {2005})},\ \Eprint
  {https://arxiv.org/abs/astro-ph/0501562} {arXiv:astro-ph/0501562}
  \BibitemShut {NoStop}%
\bibitem [{\citenamefont {Osato}\ \emph {et~al.}(2016)\citenamefont {Osato},
  \citenamefont {Sekiguchi}, \citenamefont {Shirasaki}, \citenamefont
  {Kamada},\ and\ \citenamefont {Yoshida}}]{Osato:2016ixc}%
  \BibitemOpen
  \bibfield  {author} {\bibinfo {author} {\bibfnamefont {K.}~\bibnamefont
  {Osato}}, \bibinfo {author} {\bibfnamefont {T.}~\bibnamefont {Sekiguchi}},
  \bibinfo {author} {\bibfnamefont {M.}~\bibnamefont {Shirasaki}}, \bibinfo
  {author} {\bibfnamefont {A.}~\bibnamefont {Kamada}},\ and\ \bibinfo {author}
  {\bibfnamefont {N.}~\bibnamefont {Yoshida}},\ }\bibfield  {title} {\bibinfo
  {title} {{Cosmological Constraint on the Light Gravitino Mass from CMB
  Lensing and Cosmic Shear}},\ }\href
  {https://doi.org/10.1088/1475-7516/2016/06/004} {\bibfield  {journal}
  {\bibinfo  {journal} {JCAP}\ }\textbf {\bibinfo {volume} {06}},\ \bibinfo
  {pages} {004}},\ \Eprint {https://arxiv.org/abs/1601.07386} {arXiv:1601.07386
  [astro-ph.CO]} \BibitemShut {NoStop}%
\bibitem [{\citenamefont {Chacko}\ \emph {et~al.}(2017)\citenamefont {Chacko},
  \citenamefont {Craig}, \citenamefont {Fox},\ and\ \citenamefont
  {Harnik}}]{Chacko:2016hvu}%
  \BibitemOpen
  \bibfield  {author} {\bibinfo {author} {\bibfnamefont {Z.}~\bibnamefont
  {Chacko}}, \bibinfo {author} {\bibfnamefont {N.}~\bibnamefont {Craig}},
  \bibinfo {author} {\bibfnamefont {P.~J.}\ \bibnamefont {Fox}},\ and\ \bibinfo
  {author} {\bibfnamefont {R.}~\bibnamefont {Harnik}},\ }\bibfield  {title}
  {\bibinfo {title} {{Cosmology in Mirror Twin Higgs and Neutrino Masses}},\
  }\href {https://doi.org/10.1007/JHEP07(2017)023} {\bibfield  {journal}
  {\bibinfo  {journal} {JHEP}\ }\textbf {\bibinfo {volume} {07}},\ \bibinfo
  {pages} {023}},\ \Eprint {https://arxiv.org/abs/1611.07975} {arXiv:1611.07975
  [hep-ph]} \BibitemShut {NoStop}%
\bibitem [{\citenamefont {Cyr-Racine}\ and\ \citenamefont
  {Sigurdson}(2013)}]{Cyr-Racine:2012tfp}%
  \BibitemOpen
  \bibfield  {author} {\bibinfo {author} {\bibfnamefont {F.-Y.}\ \bibnamefont
  {Cyr-Racine}}\ and\ \bibinfo {author} {\bibfnamefont {K.}~\bibnamefont
  {Sigurdson}},\ }\bibfield  {title} {\bibinfo {title} {{Cosmology of atomic
  dark matter}},\ }\href {https://doi.org/10.1103/PhysRevD.87.103515}
  {\bibfield  {journal} {\bibinfo  {journal} {Phys. Rev. D}\ }\textbf {\bibinfo
  {volume} {87}},\ \bibinfo {pages} {103515} (\bibinfo {year} {2013})},\
  \Eprint {https://arxiv.org/abs/1209.5752} {arXiv:1209.5752 [astro-ph.CO]}
  \BibitemShut {NoStop}%
\bibitem [{\citenamefont {D'Eramo}\ \emph {et~al.}(2018)\citenamefont
  {D'Eramo}, \citenamefont {Ferreira}, \citenamefont {Notari},\ and\
  \citenamefont {Bernal}}]{DEramo:2018vss}%
  \BibitemOpen
  \bibfield  {author} {\bibinfo {author} {\bibfnamefont {F.}~\bibnamefont
  {D'Eramo}}, \bibinfo {author} {\bibfnamefont {R.~Z.}\ \bibnamefont
  {Ferreira}}, \bibinfo {author} {\bibfnamefont {A.}~\bibnamefont {Notari}},\
  and\ \bibinfo {author} {\bibfnamefont {J.~L.}\ \bibnamefont {Bernal}},\
  }\bibfield  {title} {\bibinfo {title} {{Hot Axions and the $H_0$ tension}},\
  }\href {https://doi.org/10.1088/1475-7516/2018/11/014} {\bibfield  {journal}
  {\bibinfo  {journal} {JCAP}\ }\textbf {\bibinfo {volume} {11}},\ \bibinfo
  {pages} {014}},\ \Eprint {https://arxiv.org/abs/1808.07430} {arXiv:1808.07430
  [hep-ph]} \BibitemShut {NoStop}%
\bibitem [{\citenamefont {Dine}\ and\ \citenamefont
  {Nelson}(1993)}]{Dine:1993yw}%
  \BibitemOpen
  \bibfield  {author} {\bibinfo {author} {\bibfnamefont {M.}~\bibnamefont
  {Dine}}\ and\ \bibinfo {author} {\bibfnamefont {A.~E.}\ \bibnamefont
  {Nelson}},\ }\bibfield  {title} {\bibinfo {title} {{Dynamical supersymmetry
  breaking at low-energies}},\ }\href
  {https://doi.org/10.1103/PhysRevD.48.1277} {\bibfield  {journal} {\bibinfo
  {journal} {Phys. Rev. D}\ }\textbf {\bibinfo {volume} {48}},\ \bibinfo
  {pages} {1277} (\bibinfo {year} {1993})},\ \Eprint
  {https://arxiv.org/abs/hep-ph/9303230} {arXiv:hep-ph/9303230} \BibitemShut
  {NoStop}%
\bibitem [{\citenamefont {Dine}\ \emph {et~al.}(1995)\citenamefont {Dine},
  \citenamefont {Nelson},\ and\ \citenamefont {Shirman}}]{Dine:1994vc}%
  \BibitemOpen
  \bibfield  {author} {\bibinfo {author} {\bibfnamefont {M.}~\bibnamefont
  {Dine}}, \bibinfo {author} {\bibfnamefont {A.~E.}\ \bibnamefont {Nelson}},\
  and\ \bibinfo {author} {\bibfnamefont {Y.}~\bibnamefont {Shirman}},\
  }\bibfield  {title} {\bibinfo {title} {{Low-energy dynamical supersymmetry
  breaking simplified}},\ }\href {https://doi.org/10.1103/PhysRevD.51.1362}
  {\bibfield  {journal} {\bibinfo  {journal} {Phys. Rev. D}\ }\textbf {\bibinfo
  {volume} {51}},\ \bibinfo {pages} {1362} (\bibinfo {year} {1995})},\ \Eprint
  {https://arxiv.org/abs/hep-ph/9408384} {arXiv:hep-ph/9408384} \BibitemShut
  {NoStop}%
\bibitem [{\citenamefont {Giudice}\ and\ \citenamefont
  {Rattazzi}(1999)}]{Giudice:1998bp}%
  \BibitemOpen
  \bibfield  {author} {\bibinfo {author} {\bibfnamefont {G.~F.}\ \bibnamefont
  {Giudice}}\ and\ \bibinfo {author} {\bibfnamefont {R.}~\bibnamefont
  {Rattazzi}},\ }\bibfield  {title} {\bibinfo {title} {{Theories with gauge
  mediated supersymmetry breaking}},\ }\href
  {https://doi.org/10.1016/S0370-1573(99)00042-3} {\bibfield  {journal}
  {\bibinfo  {journal} {Phys. Rept.}\ }\textbf {\bibinfo {volume} {322}},\
  \bibinfo {pages} {419} (\bibinfo {year} {1999})},\ \Eprint
  {https://arxiv.org/abs/hep-ph/9801271} {arXiv:hep-ph/9801271} \BibitemShut
  {NoStop}%
\bibitem [{\citenamefont {Ichikawa}\ \emph {et~al.}(2009)\citenamefont
  {Ichikawa}, \citenamefont {Kawasaki}, \citenamefont {Nakayama}, \citenamefont
  {Sekiguchi},\ and\ \citenamefont {Takahashi}}]{Ichikawa:2009ir}%
  \BibitemOpen
  \bibfield  {author} {\bibinfo {author} {\bibfnamefont {K.}~\bibnamefont
  {Ichikawa}}, \bibinfo {author} {\bibfnamefont {M.}~\bibnamefont {Kawasaki}},
  \bibinfo {author} {\bibfnamefont {K.}~\bibnamefont {Nakayama}}, \bibinfo
  {author} {\bibfnamefont {T.}~\bibnamefont {Sekiguchi}},\ and\ \bibinfo
  {author} {\bibfnamefont {T.}~\bibnamefont {Takahashi}},\ }\bibfield  {title}
  {\bibinfo {title} {{Constraining Light Gravitino Mass from Cosmic Microwave
  Background}},\ }\href {https://doi.org/10.1088/1475-7516/2009/08/013}
  {\bibfield  {journal} {\bibinfo  {journal} {JCAP}\ }\textbf {\bibinfo
  {volume} {08}},\ \bibinfo {pages} {013}},\ \Eprint
  {https://arxiv.org/abs/0905.2237} {arXiv:0905.2237 [astro-ph.CO]}
  \BibitemShut {NoStop}%
\bibitem [{\citenamefont {Abada}\ \emph {et~al.}(2019)\citenamefont {Abada}
  \emph {et~al.}}]{FCC:2018evy}%
  \BibitemOpen
  \bibfield  {author} {\bibinfo {author} {\bibfnamefont {A.}~\bibnamefont
  {Abada}} \emph {et~al.} (\bibinfo {collaboration} {FCC}),\ }\bibfield
  {title} {\bibinfo {title} {{FCC-ee: The Lepton Collider}: {Future Circular
  Collider Conceptual Design Report Volume 2}},\ }\href
  {https://doi.org/10.1140/epjst/e2019-900045-4} {\bibfield  {journal}
  {\bibinfo  {journal} {Eur. Phys. J. ST}\ }\textbf {\bibinfo {volume} {228}},\
  \bibinfo {pages} {261} (\bibinfo {year} {2019})}\BibitemShut {NoStop}%
\bibitem [{\citenamefont {Schulte}(2017)}]{Schulte:2017qkc}%
  \BibitemOpen
  \bibfield  {author} {\bibinfo {author} {\bibfnamefont {D.}~\bibnamefont
  {Schulte}},\ }\bibfield  {title} {\bibinfo {title} {{FCC-hh Design
  Highlights}},\ }\href@noop {} {\bibfield  {journal} {\bibinfo  {journal}
  {ICFA Beam Dyn. Newslett.}\ }\textbf {\bibinfo {volume} {72}},\ \bibinfo
  {pages} {99} (\bibinfo {year} {2017})}\BibitemShut {NoStop}%
\bibitem [{\citenamefont {Hinchliffe}\ \emph {et~al.}(2015)\citenamefont
  {Hinchliffe}, \citenamefont {Kotwal}, \citenamefont {Mangano}, \citenamefont
  {Quigg},\ and\ \citenamefont {Wang}}]{Hinchliffe:2015qma}%
  \BibitemOpen
  \bibfield  {author} {\bibinfo {author} {\bibfnamefont {I.}~\bibnamefont
  {Hinchliffe}}, \bibinfo {author} {\bibfnamefont {A.}~\bibnamefont {Kotwal}},
  \bibinfo {author} {\bibfnamefont {M.~L.}\ \bibnamefont {Mangano}}, \bibinfo
  {author} {\bibfnamefont {C.}~\bibnamefont {Quigg}},\ and\ \bibinfo {author}
  {\bibfnamefont {L.-T.}\ \bibnamefont {Wang}},\ }\bibfield  {title} {\bibinfo
  {title} {{Luminosity goals for a 100-TeV pp collider}},\ }\href
  {https://doi.org/10.1142/S0217751X15440029} {\bibfield  {journal} {\bibinfo
  {journal} {Int. J. Mod. Phys. A}\ }\textbf {\bibinfo {volume} {30}},\
  \bibinfo {pages} {1544002} (\bibinfo {year} {2015})},\ \Eprint
  {https://arxiv.org/abs/1504.06108} {arXiv:1504.06108 [hep-ph]} \BibitemShut
  {NoStop}%
\bibitem [{\citenamefont {Angulo}\ \emph {et~al.}(2015)\citenamefont {Angulo},
  \citenamefont {Fasiello}, \citenamefont {Senatore},\ and\ \citenamefont
  {Vlah}}]{Angulo:2015eqa}%
  \BibitemOpen
  \bibfield  {author} {\bibinfo {author} {\bibfnamefont {R.}~\bibnamefont
  {Angulo}}, \bibinfo {author} {\bibfnamefont {M.}~\bibnamefont {Fasiello}},
  \bibinfo {author} {\bibfnamefont {L.}~\bibnamefont {Senatore}},\ and\
  \bibinfo {author} {\bibfnamefont {Z.}~\bibnamefont {Vlah}},\ }\bibfield
  {title} {\bibinfo {title} {{On the Statistics of Biased Tracers in the
  Effective Field Theory of Large Scale Structures}},\ }\href
  {https://doi.org/10.1088/1475-7516/2015/9/029} {\bibfield  {journal}
  {\bibinfo  {journal} {JCAP}\ }\textbf {\bibinfo {volume} {09}},\ \bibinfo
  {pages} {029}},\ \Eprint {https://arxiv.org/abs/1503.08826} {arXiv:1503.08826
  [astro-ph.CO]} \BibitemShut {NoStop}%
\bibitem [{\citenamefont {Senatore}\ and\ \citenamefont
  {Zaldarriaga}(2017)}]{Senatore:2017hyk}%
  \BibitemOpen
  \bibfield  {author} {\bibinfo {author} {\bibfnamefont {L.}~\bibnamefont
  {Senatore}}\ and\ \bibinfo {author} {\bibfnamefont {M.}~\bibnamefont
  {Zaldarriaga}},\ }\href@noop {} {\bibinfo {title} {{The Effective Field
  Theory of Large-Scale Structure in the presence of Massive Neutrinos}}}
  (\bibinfo {year} {2017}),\ \Eprint {https://arxiv.org/abs/1707.04698}
  {arXiv:1707.04698 [astro-ph.CO]} \BibitemShut {NoStop}%
\bibitem [{\citenamefont {Ivanov}\ \emph
  {et~al.}(2020{\natexlab{b}})\citenamefont {Ivanov}, \citenamefont
  {McDonough}, \citenamefont {Hill}, \citenamefont {Simonovi\'c}, \citenamefont
  {Toomey}, \citenamefont {Alexander},\ and\ \citenamefont
  {Zaldarriaga}}]{Ivanov:2020ril}%
  \BibitemOpen
  \bibfield  {author} {\bibinfo {author} {\bibfnamefont {M.~M.}\ \bibnamefont
  {Ivanov}}, \bibinfo {author} {\bibfnamefont {E.}~\bibnamefont {McDonough}},
  \bibinfo {author} {\bibfnamefont {J.~C.}\ \bibnamefont {Hill}}, \bibinfo
  {author} {\bibfnamefont {M.}~\bibnamefont {Simonovi\'c}}, \bibinfo {author}
  {\bibfnamefont {M.~W.}\ \bibnamefont {Toomey}}, \bibinfo {author}
  {\bibfnamefont {S.}~\bibnamefont {Alexander}},\ and\ \bibinfo {author}
  {\bibfnamefont {M.}~\bibnamefont {Zaldarriaga}},\ }\bibfield  {title}
  {\bibinfo {title} {{Constraining Early Dark Energy with Large-Scale
  Structure}},\ }\href {https://doi.org/10.1103/PhysRevD.102.103502} {\bibfield
   {journal} {\bibinfo  {journal} {Phys. Rev. D}\ }\textbf {\bibinfo {volume}
  {102}},\ \bibinfo {pages} {103502} (\bibinfo {year} {2020}{\natexlab{b}})},\
  \Eprint {https://arxiv.org/abs/2006.11235} {arXiv:2006.11235 [astro-ph.CO]}
  \BibitemShut {NoStop}%
\bibitem [{\citenamefont {Castorina}\ \emph {et~al.}(2014)\citenamefont
  {Castorina}, \citenamefont {Sefusatti}, \citenamefont {Sheth}, \citenamefont
  {Villaescusa-Navarro},\ and\ \citenamefont {Viel}}]{Castorina:2013wga}%
  \BibitemOpen
  \bibfield  {author} {\bibinfo {author} {\bibfnamefont {E.}~\bibnamefont
  {Castorina}}, \bibinfo {author} {\bibfnamefont {E.}~\bibnamefont
  {Sefusatti}}, \bibinfo {author} {\bibfnamefont {R.~K.}\ \bibnamefont
  {Sheth}}, \bibinfo {author} {\bibfnamefont {F.}~\bibnamefont
  {Villaescusa-Navarro}},\ and\ \bibinfo {author} {\bibfnamefont
  {M.}~\bibnamefont {Viel}},\ }\bibfield  {title} {\bibinfo {title} {{Cosmology
  with massive neutrinos II: on the universality of the halo mass function and
  bias}},\ }\href {https://doi.org/10.1088/1475-7516/2014/02/049} {\bibfield
  {journal} {\bibinfo  {journal} {JCAP}\ }\textbf {\bibinfo {volume} {02}},\
  \bibinfo {pages} {049}},\ \Eprint {https://arxiv.org/abs/1311.1212}
  {arXiv:1311.1212 [astro-ph.CO]} \BibitemShut {NoStop}%
\bibitem [{\citenamefont {Lazeyras}\ \emph {et~al.}(2016)\citenamefont
  {Lazeyras}, \citenamefont {Wagner}, \citenamefont {Baldauf},\ and\
  \citenamefont {Schmidt}}]{Lazeyras:2015lgp}%
  \BibitemOpen
  \bibfield  {author} {\bibinfo {author} {\bibfnamefont {T.}~\bibnamefont
  {Lazeyras}}, \bibinfo {author} {\bibfnamefont {C.}~\bibnamefont {Wagner}},
  \bibinfo {author} {\bibfnamefont {T.}~\bibnamefont {Baldauf}},\ and\ \bibinfo
  {author} {\bibfnamefont {F.}~\bibnamefont {Schmidt}},\ }\bibfield  {title}
  {\bibinfo {title} {{Precision measurement of the local bias of dark matter
  halos}},\ }\href {https://doi.org/10.1088/1475-7516/2016/02/018} {\bibfield
  {journal} {\bibinfo  {journal} {JCAP}\ }\textbf {\bibinfo {volume} {02}},\
  \bibinfo {pages} {018}},\ \Eprint {https://arxiv.org/abs/1511.01096}
  {arXiv:1511.01096 [astro-ph.CO]} \BibitemShut {NoStop}%
\bibitem [{\citenamefont {Xu}\ \emph {et~al.}(2021)\citenamefont {Xu},
  \citenamefont {DePorzio}, \citenamefont {Mu\~noz},\ and\ \citenamefont
  {Dvorkin}}]{Xu:2020fyg}%
  \BibitemOpen
  \bibfield  {author} {\bibinfo {author} {\bibfnamefont {W.~L.}\ \bibnamefont
  {Xu}}, \bibinfo {author} {\bibfnamefont {N.}~\bibnamefont {DePorzio}},
  \bibinfo {author} {\bibfnamefont {J.~B.}\ \bibnamefont {Mu\~noz}},\ and\
  \bibinfo {author} {\bibfnamefont {C.}~\bibnamefont {Dvorkin}},\ }\bibfield
  {title} {\bibinfo {title} {{Accurately Weighing Neutrinos with Cosmological
  Surveys}},\ }\href {https://doi.org/10.1103/PhysRevD.103.023503} {\bibfield
  {journal} {\bibinfo  {journal} {Phys. Rev. D}\ }\textbf {\bibinfo {volume}
  {103}},\ \bibinfo {pages} {023503} (\bibinfo {year} {2021})},\ \Eprint
  {https://arxiv.org/abs/2006.09395} {arXiv:2006.09395 [astro-ph.CO]}
  \BibitemShut {NoStop}%
\bibitem [{\citenamefont {Vagnozzi}\ \emph {et~al.}(2018)\citenamefont
  {Vagnozzi}, \citenamefont {Brinckmann}, \citenamefont {Archidiacono},
  \citenamefont {Freese}, \citenamefont {Gerbino}, \citenamefont
  {Lesgourgues},\ and\ \citenamefont {Sprenger}}]{Vagnozzi:2018pwo}%
  \BibitemOpen
  \bibfield  {author} {\bibinfo {author} {\bibfnamefont {S.}~\bibnamefont
  {Vagnozzi}}, \bibinfo {author} {\bibfnamefont {T.}~\bibnamefont
  {Brinckmann}}, \bibinfo {author} {\bibfnamefont {M.}~\bibnamefont
  {Archidiacono}}, \bibinfo {author} {\bibfnamefont {K.}~\bibnamefont
  {Freese}}, \bibinfo {author} {\bibfnamefont {M.}~\bibnamefont {Gerbino}},
  \bibinfo {author} {\bibfnamefont {J.}~\bibnamefont {Lesgourgues}},\ and\
  \bibinfo {author} {\bibfnamefont {T.}~\bibnamefont {Sprenger}},\ }\bibfield
  {title} {\bibinfo {title} {{Bias due to neutrinos must not uncorrect'd go}},\
  }\href {https://doi.org/10.1088/1475-7516/2018/09/001} {\bibfield  {journal}
  {\bibinfo  {journal} {JCAP}\ }\textbf {\bibinfo {volume} {09}},\ \bibinfo
  {pages} {001}},\ \Eprint {https://arxiv.org/abs/1807.04672} {arXiv:1807.04672
  [astro-ph.CO]} \BibitemShut {NoStop}%
\bibitem [{\citenamefont {Chiang}\ \emph {et~al.}(2019)\citenamefont {Chiang},
  \citenamefont {LoVerde},\ and\ \citenamefont
  {Villaescusa-Navarro}}]{Chiang:2018laa}%
  \BibitemOpen
  \bibfield  {author} {\bibinfo {author} {\bibfnamefont {C.-T.}\ \bibnamefont
  {Chiang}}, \bibinfo {author} {\bibfnamefont {M.}~\bibnamefont {LoVerde}},\
  and\ \bibinfo {author} {\bibfnamefont {F.}~\bibnamefont
  {Villaescusa-Navarro}},\ }\bibfield  {title} {\bibinfo {title} {{First
  detection of scale-dependent linear halo bias in $N$-body simulations with
  massive neutrinos}},\ }\href {https://doi.org/10.1103/PhysRevLett.122.041302}
  {\bibfield  {journal} {\bibinfo  {journal} {Phys. Rev. Lett.}\ }\textbf
  {\bibinfo {volume} {122}},\ \bibinfo {pages} {041302} (\bibinfo {year}
  {2019})},\ \Eprint {https://arxiv.org/abs/1811.12412} {arXiv:1811.12412
  [astro-ph.CO]} \BibitemShut {NoStop}%
\bibitem [{\citenamefont {Vagnozzi}\ \emph {et~al.}(2017)\citenamefont
  {Vagnozzi}, \citenamefont {Giusarma}, \citenamefont {Mena}, \citenamefont
  {Freese}, \citenamefont {Gerbino}, \citenamefont {Ho},\ and\ \citenamefont
  {Lattanzi}}]{Vagnozzi:2017ovm}%
  \BibitemOpen
  \bibfield  {author} {\bibinfo {author} {\bibfnamefont {S.}~\bibnamefont
  {Vagnozzi}}, \bibinfo {author} {\bibfnamefont {E.}~\bibnamefont {Giusarma}},
  \bibinfo {author} {\bibfnamefont {O.}~\bibnamefont {Mena}}, \bibinfo {author}
  {\bibfnamefont {K.}~\bibnamefont {Freese}}, \bibinfo {author} {\bibfnamefont
  {M.}~\bibnamefont {Gerbino}}, \bibinfo {author} {\bibfnamefont
  {S.}~\bibnamefont {Ho}},\ and\ \bibinfo {author} {\bibfnamefont
  {M.}~\bibnamefont {Lattanzi}},\ }\bibfield  {title} {\bibinfo {title}
  {{Unveiling $\nu$ secrets with cosmological data: neutrino masses and mass
  hierarchy}},\ }\href {https://doi.org/10.1103/PhysRevD.96.123503} {\bibfield
  {journal} {\bibinfo  {journal} {Phys. Rev. D}\ }\textbf {\bibinfo {volume}
  {96}},\ \bibinfo {pages} {123503} (\bibinfo {year} {2017})},\ \Eprint
  {https://arxiv.org/abs/1701.08172} {arXiv:1701.08172 [astro-ph.CO]}
  \BibitemShut {NoStop}%
\bibitem [{\citenamefont {Abbott}\ \emph
  {et~al.}(2021{\natexlab{b}})\citenamefont {Abbott} \emph
  {et~al.}}]{DES:2021wwk}%
  \BibitemOpen
  \bibfield  {author} {\bibinfo {author} {\bibfnamefont {T.~M.~C.}\
  \bibnamefont {Abbott}} \emph {et~al.} (\bibinfo {collaboration} {DES}),\
  }\href@noop {} {\bibinfo {title} {{Dark Energy Survey Year 3 Results:
  Cosmological Constraints from Galaxy Clustering and Weak Lensing}}} (\bibinfo
  {year} {2021}{\natexlab{b}}),\ \Eprint {https://arxiv.org/abs/2105.13549}
  {arXiv:2105.13549 [astro-ph.CO]} \BibitemShut {NoStop}%
\bibitem [{\citenamefont {Hildebrandt}\ \emph {et~al.}(2020)\citenamefont
  {Hildebrandt} \emph {et~al.}}]{Hildebrandt:2018yau}%
  \BibitemOpen
  \bibfield  {author} {\bibinfo {author} {\bibfnamefont {H.}~\bibnamefont
  {Hildebrandt}} \emph {et~al.},\ }\bibfield  {title} {\bibinfo {title}
  {{KiDS+VIKING-450: Cosmic shear tomography with optical and infrared data}},\
  }\href {https://doi.org/10.1051/0004-6361/201834878} {\bibfield  {journal}
  {\bibinfo  {journal} {Astron. Astrophys.}\ }\textbf {\bibinfo {volume}
  {633}},\ \bibinfo {pages} {A69} (\bibinfo {year} {2020})},\ \Eprint
  {https://arxiv.org/abs/1812.06076} {arXiv:1812.06076 [astro-ph.CO]}
  \BibitemShut {NoStop}%
\end{thebibliography}%

\end{document}